\documentclass[journal, onecolumn]{IEEEtran}

\usepackage[pdftex]{graphicx}
\usepackage{amsthm}
\usepackage{amsmath}
\usepackage{CJK}
\usepackage{hyperref}
\usepackage{mathtools}

\title{Large Alphabet Compression and Predictive Distributions through Poissonization and Tilting}

\author{Xiao~Yang,~\IEEEmembership{Student member,~IEEE}
            Andrew R.~Barron,~\IEEEmembership{Fellow,~IEEE}%
\thanks{Part of the paper was presented at The Sixth Workshop on Information Theoretic Methods in Science and 
             Engineering, 26-29 August 2013, Tokyo, Japan}}

\begin{document}

% make the title area
\maketitle
\IEEEpeerreviewmaketitle

\begin{abstract}
This paper introduces a convenient strategy for coding and predicting sequences of independent, identically distributed random variables generated from a large alphabet of size $m$. In particular, the size of the sample is allowed to be variable. The employment of a Poisson model and tilting method simplifies the implementation and analysis through independence. The resulting strategy is optimal within the class of distributions satisfying a moment condition, and is close to optimal for the class of all i.i.d distributions on strings of a given length. Moreover, the method can be used to code and predict strings with a condition on the tail of the ordered counts. It can also be applied to distributions in an envelope class. 
\end{abstract}


\section{Introduction}
\IEEEPARstart{L}{arge} alphabet compression and prediction problems concern understanding the probabilistic scheme of a huge number of possible outcomes. In many cases the ordered probability of individual outcomes displays a quickly falling shape, with a small number of outcomes happening most often. An example is Chinese characters. A recent published dictionary contains 85568 Chinese characters in total \cite{chinese}, but the number of frequent characters is considerably smaller. Here we consider an i.i.d model for this problem. Despite the possible dependence among the symbols in the alphabet as in language, it serves as a start and can be extended to models taking dependence into account. 

Previous theoretical analysis usually assumes the length of a message is known in advance when it is coded. This is not always true in practice. Serialization writers do not know how many words a novel contains exactly before he finishes the last sentence. Nevertheless, given a limited time or space, one could possibly guess how many words on average can be accommodated.

Suppose a string of random variables $\underline{X} = (X_1,\ldots,X_N)$ is generated independently from a discrete alphabet $\mathcal A$ of size $m$. We allow the string length $N$ to be variable. A special case is when $N$ is given as a fixed number, or it can be random. In either case, $\underline X$ is a member of the set $\mathcal X^*$ of all finite length srtings
\begin{eqnarray}
\mathcal X^* &=& \bigcup_{n=0}^\infty \mathcal X^n  \nonumber \\
                     &=& \bigcup_{n=0}^\infty \{x^n\!=\!(x_1,\ldots,x_n): x_i \in \mathcal A, i=1,\ldots,n \}. \nonumber
\end{eqnarray}
Our goal is to code/predict the string $\underline X$. Note that the length $N$ is determined by the string. There will be an agreed upon distribution of $N$, perhaps Poisson or deterministic.

Now suppose given $N$, each random variable $X_i$ is generated independently according to a probability mass function in a parametric family $\mathcal P_{\Theta}=\{P_{\underline \theta}(x): \mathbf{\underline \theta} \in \Theta \subset R^m \}$ on $\mathcal A$. Thus 
$$ P_{\underline \theta} (X_1,\ldots,X_N | N=n) = \prod_{i=1}^n P_{\underline \theta} (X_i) $$
for $n=1,2,\ldots$ Of particular interest is the class of all distributions with $P_{\underline \theta}(j)=\theta_j$ parameterized by the simplex $\Theta=\{\underline \theta\!=\!(\theta_1,\ldots,\theta_m)\!: \theta_j\!\geq\!0, \sum_{j=1}^m \theta_j\!=\!1, j\!=\!1,\ldots,m \}$.

Let $\underline N = (N_1,\ldots, N_m)$ denote the vector of counts for symbol $1,\ldots,m$. The observed sample size $N$ is the sum of the counts $N=\sum_{j=1}^m N_j$. Both $P_{\underline \theta} (\underline X)$ and $P_{\underline \theta} (\underline X | N=n)$ have factorizations based on the distribution of the counts
\begin{eqnarray}
P_{\underline \theta} (\underline X | N\!=\!n ) = P(\underline X | \underline N)\, P_{\underline \theta}(\underline N | N\!=\!n), \nonumber
\end{eqnarray}
and
\begin{eqnarray}
P_{\underline \theta} (\underline X) &=& P(\underline X | \underline N)\, P_{\underline \theta}(\underline N). \nonumber
\end{eqnarray}
The first factor of the two equations is the uniform distribution on the set of strings with given counts, which does not depend on $\underline \theta$. The vector of counts $\underline N$ forms a sufficient statistic for $\underline \theta$. Modeling the distribution of the counts is essential for forming codes and predictions. In the particular case of all i.i.d. distributions parameterized by the simplex, the distribution $P_{\underline \theta} (\underline N | N=n)$ is the $multinomial(n, \underline \theta)$ distribution. 

In the above, there is a need for a distribution of the total count $N$. Of particular interest is the case that the total count is taken to be $Poisson$, because then the resulting distribution of individual counts makes them independent. 

Accordingly, we give particular attention to the target family $\mathcal P_{\Lambda}^m = \{ P_{\underline \lambda} (\underline N)\!: \lambda_j\!\geq\!0, \,j\!=\!1,\ldots,m \}$, in which $P_{\underline \lambda} (\underline N)$ is the product of $Poisson(\lambda_j)$ distribution for $N_j, j\!=\!1,\ldots,m$. It makes the total count $N \sim Poisson(\lambda_{sum})$ with $\lambda_{sum}=\sum_{j=1}^m \lambda_j$ and yields the $multinomial(n,\underline \theta)$ distribution by conditioning on $N=n$, where $\theta_j=\lambda_j/\lambda_{sum}$. And the induced distribution on $\underline X$ is
$$ P_{\underline \lambda} (\underline X) = P (\underline X | \underline N) P_{\underline \lambda} (\underline N). $$

The task of coding a string is equivalent to providing a probabilistic scheme. A coder $Q$ for the string is also a (sub)probability distribution on $\mathcal X^*$ which assigns a probability $Q(\underline X)$ to each string $\underline X$ and produces a binary string of length $\log 1/Q(\underline X)$ (we do not worry about the integer constraint). Ideally the true probability distribution $P_{\lambda} (\underline X)$ could be used if $\underline \lambda$ were known, as it produces no extra bits for coding purpose. The {\em regret} induced by using $Q$ instead of $P_{\underline \lambda}$ is
$$ R(Q, P_{\underline \lambda}, \underline X) = \log \frac{1}{Q(\underline X)} - \log \frac{1}{P_{\underline \lambda} (\underline X)}, $$
where $\log$ is logarithm base $2$. Likewise, the {\em expected regret} is
$$ r (Q, P_{\underline \lambda}) = \mathbf{E}_{P_{\underline \lambda}} \left( \log \frac{1}{Q(\underline{X})} - \frac{1}{P_{\underline \lambda} (\underline{X})} \right). $$
In universal coding the expected regret is also called the {\em redundancy}.

Here we can construct $Q$ by choosing a probability distribution for the counts and then use the uniform distribution for the distribution of strings given the counts, written as $P_{unif}$. That is
\begin{equation}
Q ( \underline X ) = P_{unif} ( \underline X | \underline N ) Q ( \underline N ). \nonumber
\end{equation}
Then the regret becomes the $\log$ ratio of the counts probability
\begin{eqnarray}
R(Q, P_{\underline \lambda}, \underline X) &=& \log \frac{P_{\underline \lambda} (\underline N)}{Q(\underline N)} \nonumber \\
                                                                     &=& R(Q, P_{\underline \lambda}, \underline N). \nonumber
\end{eqnarray}
And the redundancy becomes
$$ r (Q, P_{\underline \lambda}) = \mathbf{E}_{P_{\underline \lambda}} \log \frac{P_{\underline \lambda} (\underline N)}{Q(\underline N)}. $$

In the pointwise regret story, the set of codelengths $\textstyle \log(1/P_{\underline \lambda}(\underline X))$ provides a standard with which our coder is to be compared. Given the family $\mathcal P_{\Lambda}^m$, consider the best candidate with hindsight $P_{\underline {\hat \lambda}} (\underline X)$, which achieves the maximum value, $P_{\underline {\hat \lambda}}(\underline X) = \max_{\underline \lambda \in \Lambda} (P_{\underline \lambda} (\underline X))$ (corresponding to $\min_{\underline \lambda \in \Lambda} \log(1/P_{\underline \lambda}(\underline X))$), where $\underline {\hat \lambda}$ is the maximum likelihood estimator of $\underline \lambda$, and compare it to our strategy $Q(\underline X)$. The maximization is equivalent to maximizing $\underline \lambda$ for the count probability, as the uniform distribution dose not depend on $\lambda$, i.e.
\begin{eqnarray}
\max_{\underline \lambda \in \Lambda} (P_{\underline \lambda}(\underline X))\!&=&\!P_{unif}( \underline X | \underline N )\, \max_{\underline \lambda \in \Lambda} P_{\underline \lambda} ( \underline N )  \nonumber \\ 
                                                                                                                           \!&=&\!P_{unif} ( \underline X | \underline N )\, P_{\underline {\hat \lambda}} (\underline N).\nonumber
\end{eqnarray}

Then the problem becomes: given the family $\mathcal P_{\Lambda}^m$, how to choose $Q$ to minimize the maximized regret
$$\min_{Q} \max_{\underline X} R(Q, P_{\underline {\hat \lambda}}, \underline{X}) = \min_{Q} \max_{\underline N}\log \frac{ P_{\underline {\hat \lambda}} (\underline N)}{Q(\underline N)},$$
or the redundancy,
$$ \min_{Q} \max_{P_{\underline \lambda} \in \mathcal P_{\Lambda}^m} r (Q, P_{\underline \lambda}) = \min_{Q} \max_{P_{\underline \lambda} \in \mathcal P_{\Lambda}^m} \mathbf{E}_{P_{\underline \lambda}} \log \frac{P_{\underline \lambda} (\underline N)}{Q(\underline N)}. $$

For the regret, the maximum can be restricted to a set of counts instead of the whole space. A traditional choice being $S_{m,n}\!=\{(N_1,\ldots,N_m)\!:\!\sum_{j=1}^m\!N_j\!=\!n, N_j \geq 0,\, j\!=\!1,\ldots,m\}$ associated with a given sample size $n$, in which case the minimax regret is
$$ \min_{Q} \max_{\underline N \in S_{m,n}}\log \frac{P_{\hat \lambda} (\underline N)}{Q(\underline N)}.$$

As is familiar in universal coding \cite{shtarkov87}\cite{xie97}, the normalized maximum likelihood (NML) distribution 
$$ Q_{nml} (\underline N) = \frac{ P_{\hat {\underline \lambda}} (\underline N)}{C(S_{m,n})} $$
is the unique pointwise minimax strategy when $ C(S_{m,n}) = \sum_{\underline N \in S_{m,n}} P_{\hat {\underline \lambda}} (\underline N) $ is finite, and $\log C(S_{m,n})$ is the minimax value. When $m$ is large, the NML distribution can be unwieldy to compute for compression or prediction. Instead we will introduce a slightly suboptimal coding distribution that makes the counts independent and show that it is nearly optimal for every $S_{m,n'}$ with $n'$ not too different from a target $n$. Indeed, we advocate that our simple coding distribution is preferable to use computationally when $m$ is large even if the sample size $n$ were known in advance.

To produce our desired coding distribution we make use of two basic principles. One is that the multinomial family of distributions on counts matches the conditional distribution of $N_1,\ldots,N_m$ given the sum $N$ when unconditionally the counts are independent Poisson. Another is the information theory principle \cite{csiszar75}\cite{csiszar84}\cite{campenhout81} that the conditional distribution given a sum (or average) of a large number of independent random variables is approximately a product of distributions, each of which is the one closest in relative entropy to the unconditional distribution subject to an expectation constraint. This minimum relative entropy distribution is an exponential tilting of the unconditional distribution. 

In the Poisson family with distribution $\textstyle \lambda_j^{N_j} e^{-\lambda_j} / N_j!$, exponential tilting (multiplying by the factor $\textstyle e^{-aN_j}$) preserves the Poisson family (with the parameter scaled to $\textstyle \lambda_j e^{-a}$). Those distributions continue to correspond to the multinomial distribution (with parameters $\theta_j = \lambda_j/\lambda_{sum}$) when conditioning on the sum of counts $N$. A particular choice of $a = \ln (\lambda_{sum}/N)$ provides the product of Poisson distributions closest to the multinomial in regret. Here for universal coding, we find the tilting of individual maximized likelihood that makes the product of such closest to the Shtarkov's NML distribution. This greatly simplifies the task of approximate optimal universal compression and the analysis of its regret.

Indeed, applying the maximum likelihood step to a Poisson count $k$ produces a maximized likelihood value of $\textstyle M(k)= k^k e^{-k} /k!$. We call this maximized likelihood the Stirling ratio, as it is the quantity that Stirling's approximation shows near $\textstyle (2 \pi k)^{-1/2}$ for $k$ not small. We find that this $M(k)$ plays a distinguished role in universal large alphabet compression, even for sequences with small counts k. This measure $M$ has a product extension to counts $N_1,N_2,\ldots,N_m$,
$$ M(\underline N ) = M(N_1)M(N_2) \cdots M(N_m). $$
Although $M$ has an infinite sum by itself, it is normalizable when tilted for every positive $a$. The tilted Stirling ratio distribution is
\begin{equation}
P_a (N_j) = \frac{N_j^{N_j} e^{-N_j}}{N_j!} \frac{e^{-aN_j}}{C_a}, \label{p.a}
\end{equation}
with the normalizer $\textstyle C_a = \sum_{k=0}^\infty k^{k} e^{-(1+a)k} / k!$. 

The coding distribution we propose and analyze is simply the product of those tilted one-dimensional maximized Poisson likelihood distributions for a value of $a$ we will specify later
$$ Q_a(\underline N) = P_a^m(\underline N) = P_a(N_1) \cdots P_a(N_m).  $$ 
By allowing description of all possible counts $N_j \geq 0$, $j=1,\ldots,m$, our codelength will be greater for some strings than codelengths designed for the case of a given sum $N=n$. Nevertheless, with $N$ distributed $Poisson (n)$, the probability of the outcome $N=n$ is approximately $P ( N=n ) \approx 1/\sqrt{2\pi n}$. So the allowance of description of $N$ (not just $N_1,\ldots,N_m$ given $N$) adds $\log 1/P( N= n)$ which is approximately $\textstyle \frac{1}{2} \log 2\pi n$ bits to the description length beyond that which would have been ideal $\log 1/Q_a( N_1,\ldots,N_m | N=n )$ if $N=n$ were known. This ideal codelength constructed from the tilted maximized Poisson, when conditioning on $n$, matches the Shtarkov's normalized maximum likelihood based on the multinomial.

For small alphabet with $m << n$, the minimax regret is about $\textstyle \frac{1}{2} \log n$ bits per free parameter (a total of $\textstyle \frac{m-1}{2} \log n$ + constant); and for large alphabet when $m \sim n$ and $ n=o(m) $, the minimax regret is about $ O(n) $ and $ \textstyle n \log \frac{m}{n}$ respectively \cite{shtarkov87}\cite{xie97}\cite{orlitsky03}\cite{szpankowski11}. The additional $\textstyle \frac{1}{2} \log n$ bits is a small price to pay for the sake of gaining the coding simplification and additional flexibility.

\begin{figure}
\begin{center}
\includegraphics[width=2.5in]{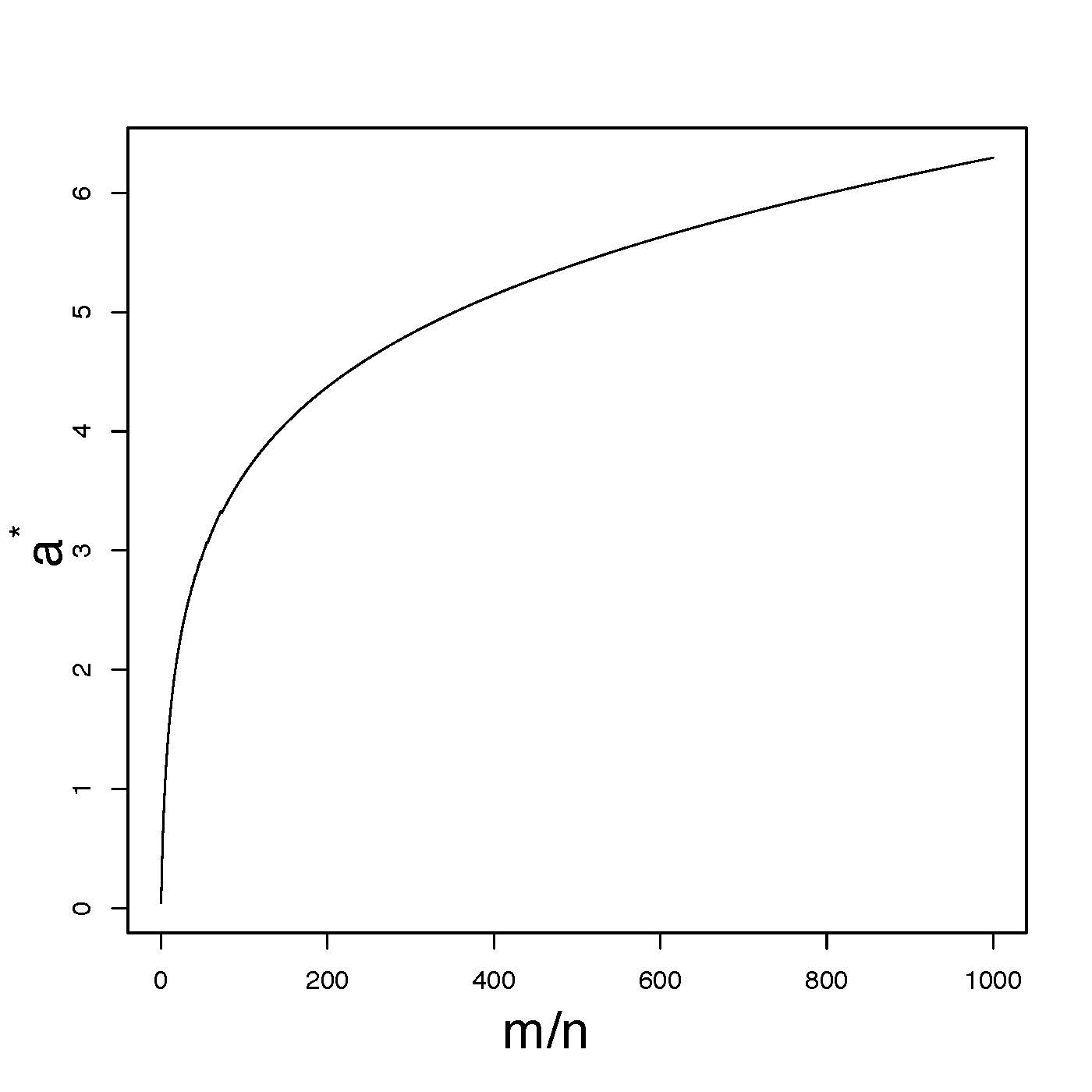}
\caption{Relationship between $a^*$ and $\frac{m}{n}$.}
\label{a_opt}
\end{center}
\end{figure}

If it is known that the total count is $n$, then the regret is a simple function of $n$ and the normalizer $C_a$. The choice of the tilting parameter $a^*$ given by the moment condition $ \mathbf{E}_{Q_a} \sum_{j=1}^m N_j = n $ minimizes the regret over all positive $a$. This arises by differentiation because $\textstyle \frac{\partial }{\partial a} \log C_a$ is equal to the given moment. Moreover, $a^*$ depends only on the ratio between the size of the alphabet and the total count $ \textstyle m/n $. Fig. \ref{a_opt} displays $a^*$ as a function of $\textstyle m/n$ solved numerically. Given an alphabet with $m$ symbols and a string generated from it of length $n$, one can look at the plot and find the $a^*$ desired according to the $\textstyle m/n$ given, and then use the $a^*$ to code the data. 

If, however, the total count $N$ is not given, then the regret depends on $N$. We use a mixture of $a$ to account for the lack of knowledge in advance, and details are discussed in section \ref{no.n}.

When $a$ is small, the tilting of the maximized Poisson likelihood distributions does not have much effect except in the tail of the distribution.  Over most of the range of count values $k$ it follows the approximate power-law $\textstyle 1/k^{1/2}$ as we have indicated.  Power-laws have been studied for count distributions and are shown to be related to Zipf's law for the sorted counts \cite{zipf}. Our use of a distribution close to a power-law is not because a power-law is assumed to govern the data, but rather because of its near optimum regret properties within suitable sets of counts, demonstrated here for the class of all Poisson count distributions, from which we obtain also its near optimality for the class of all Multinomial distributions on counts.

Shtarkov studied the universal data compression problem and identified the exact pointwise minimax strategy \cite{shtarkov87}. He showed the asymptotic minimax lower bound for the regret is $\textstyle \frac{m-1}{2} \log n + O(1)$, in which the parameter set $\Theta$ is the $m-1$ dimensional simplex of all probability vectors on an alphabet of size $m$. However, this strategy cannot be easily implemented for prediction or compression \cite{shtarkov87}, because of the computational inconvenience of computing the normalizing constant, and because of the difficulty in computing the successive conditionals required for implementation (by arithmetic coding). Let $m^*$ be the number of different symbols that appear in a sequence. Shtarkov\cite{shtarkov95} also pointed out that when $m$ is large, it typical that $m^*$ is much less than $m$, and the regret depends mainly on $m^*$ rather than $m$. Xie and Barron\cite{xie97}\cite{xie00} gave an asymptotic minimax strategy for coding under both the expected and pointwise regret for fixed size alphabet, which is formulated by a modification of the mixture density using Jeffery's prior. The asymptotic value of both the redundancy and the regret are of the form $\textstyle \frac{m-1}{2} \log n + C_m + o(1)$, where $C_m$ is a constant depending on $m$. Orlitsky and Santhanam\cite{orlitsky04.2} considered the problem in a large alphabet setting in which the number of symbols $m$ is much larger than the sequence length $n$ or even infinite. They found the main terms in the minimax regret for $m=o(n)$, $m \sim n$ and $n=o(m)$ cases take the forms $\textstyle \frac{m-1}{2} \log \frac{n}{m}$, $O(m)$ and $n \log \frac{m}{n}$ respectively. Szpankowski and Weinberger\cite{szpankowski11} provided more precise asymptotics in these settings. They also calculated the minimax regret of a source model in which some symbol probabilities are fixed. Boucheron, Garivier and Gassiat\cite{boucheron08} focused on countably infinite alphabets with an envelope condition; they used an adapted strategy and gave upper and lower bounds for pointwise minimax regret. Later on Bontemps and Gassiat\cite{bontemps11} worked on exponentially decreasing envelope class and provided a minimax strategy and the corresponding regret.

In this paper, we introduce a straightforward and easy to implement method for large alphabet coding. The purpose is three-fold: first, by allowing the sample size to be variable, we are considering a larger class of distributions. This is a more realistic and less restrictive assumption than presuming a particular length. But the method can also be used for fixed sample size coding and prediction. 

Second, it unveils an information geometry of three key distributions/measures in the problem: the unnormalized maximum Poisson likelihood measure $M$ of the counts, the conditional distribution $M_{cond}$ of $M$ given the total count equals $n$, which matches Shtarkov's normalized maximum multinomial likelihood distribution, and a tilted distribution $Q_a$, with the tilting parameter $a$ chosen to make the expected total count equal to $n$. This tilted distribution $Q_a$ minimizes the relative entropy from the original measure $M$ within the class $\mathcal C$ of distributions with the moment condition $E[N]=n$. Hence, $Q_a$ is the information projection of $M$ onto $\mathcal C$. Moreover, since $M_{cond}$ is also in $\mathcal C$, the Pythagorean-like equality holds \cite{kullback59}\cite{csiszar75}, i.e.
$$ D(M_{cond} || M) = D(M_{cond} || Q_a) + D(Q_a || M). $$
The case of a tilted distribution (the information projection) as an approximating conditional distribution is investigated in  \cite{campenhout81} and \cite{csiszar84}. A difference here is that our unconditional measure $M$ is not normalizable.

Thirdly, the strategy designed through an independent Poisson model and tilting is much easier to analyze and compute as compared to the strategies based on multinomials. The convenience is gained through independence. To actually apply this two pass code, one could first describe the independent counts $N_1,\ldots, N_m$, for instance by arithmetic coding using $P_a(N_j)$, and then describe $X_1,\ldots,X_n$ given the count vector, by arithmetic coding using the sequence of conditional distributions for $X_{i+1}$ given both $X_1,\ldots,X_i$ and all the counts (which is the sampling without replacement distribution, proportional to the counts of what remains after step $i$). 

This paper is organized in the following way. Section II introduces the model. Section III provides general results and outlines the proof, and Section IV gives simulated and real data examples. Details of proof are left in the appendix.

\section{The Poisson Model}
\label{model}

A Poisson model fits well into this problem. We have for each $j=1,\ldots,m$,
$$ N_j \sim Poisson(\lambda_j), $$
independently, and $N$ also has a Poisson distribution 
$$ N \sim Poisson(\lambda_{sum}), $$
where $\lambda_{sum}=\sum_{j=1}^m \lambda_j$. Write $\underline \lambda = (\lambda_1, \ldots, \lambda_m)$, we have
\begin{equation}
P_{\underline \lambda} (\underline X) = P_{unif} ( \underline X | \underline N ) \prod_{j=1}^m P_{\lambda_j}(N_j). \nonumber
\end{equation}

We know that the the MLE for each $\lambda_j$ is $\hat \lambda_j = N_j$, and the first term is a uniform distribution which does not depend on $\underline \lambda$. So
\begin{equation}
P_{\underline {\hat \lambda}}( \underline X ) = P_{unif} ( \underline X | \underline N ) \prod_{j=1}^m M(N_j). \nonumber
\end{equation}  
where $M(k)=k^k e^{-k}/k!$, $k=1,2,\ldots$ (as given in the introduction) is the unnormalized maximized likelihood $\textstyle M(N_j)=\max_{\lambda_j} P_{\lambda_j} (N_j)$.

If we use a distribution $Q(\underline N)$ to code the counts, then the regret is
\begin{equation}
\log \frac{ P_{\hat{\underline \lambda}}( \underline X ) }{ P( \underline X | \underline N ) Q (\underline N) } = \log \frac{\prod_{j=1}^m M(N_j) }{Q(\underline N)}. \nonumber
\end{equation}

And the redundancy is
\begin{equation}
\mathbf{E}_{P_{\underline \lambda}} \log \frac{ P( \underline X | \underline \lambda ) }{ P( \underline X | \underline N ) Q (\underline N) } = \mathbf{E}_{P_{\underline \lambda}} \log \frac{P( \underline N  | \underline \lambda ) }{Q(\underline N)}. \nonumber
\end{equation}

This method can also be applied to fixed total count scenario, which corresponds to the multinomial coding and prediction problem. Suppose $N=n$ is given, the Poisson model, when conditioned on $N=n$, indeed reduces to the i.i.d sampling model
$$ P_{\underline \lambda} ( X_1, \ldots , X_N | N=n) =  P_{\underline \theta} ( X_1, \ldots , X_n).$$

The right hand side is a discrete memoryless source distribution (i.i.d. $P_{\underline \theta}$) with probability specified by 
$P_{\underline \theta} (j) = \theta_j$, for $j=1,\ldots,m$. Note that a sequence $X_1,\ldots,X_N$ with counts $N_1,\ldots,N_m$ of total $N=n$ satisfies
\begin{eqnarray}
&& P_{\underline \lambda} (X_1,\ldots,X_N | N=n) \nonumber \\
&=& \frac{P_{\underline \lambda} (X_1,\ldots,X_n)}{P_{\lambda_{sum}} (N=n)} \nonumber \\
&=& \frac{P_{unif} (X_1,\!\ldots\!,X_n | N_1,\!\ldots\!,N_m) P_{\underline \lambda} (N_1,\!\ldots\!,N_m)}{P_{\lambda_{sum}} (N\!=\!n)}. \nonumber
\end{eqnarray}

The question left is still how to model the counts. The maximized likelihood (the same target as used by Shtarkov) is thus expressible as
\begin{eqnarray}
\nonumber
&& P_{\hat{\underline \lambda}} ( X_1,\ldots,X_N | N=n )  \nonumber \\
&=& \frac{P_{unif} ( X_1,\ldots,X_n | N_1,\ldots,N_m ) \prod_{j=1}^m M(N_j)}{P_{\hat {\lambda}_{sum}} ( N=n)}. \nonumber 
\end{eqnarray}  

Now again if we use $Q(N_1, \ldots, N_m)$ to code the counts, then the regret is
\begin{eqnarray}
\nonumber
&& \log \frac{ P_{ \hat{\underline \lambda}} ( X_1,\!\ldots\!,X_N | N=n ) }{ P_{unif} ( X_1,\!\ldots\!,X_n | N_1,\!\ldots\!,N_m ) Q (N_1,\!\ldots\!, N_m) }  \nonumber \\
&=&  \log \frac{\prod_{j=1}^m M(N_j)}{ P_{\hat \lambda_{sum}} ( N=n ) Q (N_1, \ldots, N_m)} \nonumber \\
&\simeq& \frac{1}{2} \log 2\pi n + \log \frac{\prod_{j=1}^m M(N_j)}{Q (N_1, \ldots, N_m)}  \label{N.given}
\end{eqnarray}
Here $\hat \lambda_{sum}=n$, hence the term $\textstyle \frac{1}{2} \log 2\pi n$ is Stirling's approximation of $\textstyle \log 1/P_{\hat{\lambda}_{sum}} ( N=n )$. The  $\textstyle \frac{1}{2} \log 2\pi n$ arises because here $Q$ includes description of the total $N$ while the more restrictive target regards it as given.

\section{Results}
\subsection{Regret}
We start by looking at the performance of using independent tilted Stirling ratio distributions as a coding strategy, by examining the resulting regret.

Let $S$ be any set of counts, then the maximized regret of using $Q$ as a coding strategy given a class $\mathcal P$ of distributions when the vector of counts is restricted to $S$ is
$$ R(Q, \mathcal P, S) = \max_{\underline N \in S} \log \frac{\max_{P \in \mathcal P} P(\underline N)}{Q(\underline N)}. $$

\newtheorem{thm}{Theorem} 

\begin{thm}
\label{R}
Let $\textstyle P_a$ be the distribution specified in equation (\ref{p.a}) (Poisson maximized likelihood, tilted and normalized). The regret of using a product of tilted distributions $Q_a = \otimes_{j=1}^m P_a$ for a given vector of counts $\underline N = (N_1,\ldots,N_m)$ is
$$ R \left( Q_a, \mathcal P_{\Lambda}^m, \underline N \right) = a N \log e + m \log C_a.  $$
Let $S_{m,n}$ be the set of count vectors with total count $n$ be defined as before, then
\begin{equation}
R \left(Q_a, \mathcal P_{\Lambda}^m, S_{m,n} \right) = a n \log e + m \log C_a. \label{R.n}
\end{equation}

Let $a^*$ be the choice of $a$ satisfying the following moment condition
\begin{equation}
\mathbf E_{P_a} \sum_{j=1}^m N_j = m\, \mathbf E_{P_a} N_1 = n. \label{m.a}
\end{equation}
Then $a^*$ is the minimizer of the regret in expression (\ref{R.n}). Write $R_{m,n} = \min_{a} R(Q_a, \mathcal P_{\Lambda}^m, S_{m,n})$.

When $m=o(n)$, the $R_{m,n}$ is near $\textstyle \frac{m}{2} \log \frac{ne}{m}$ in the following sense.
\begin{eqnarray}
-d_1 \frac{m}{2} \log e &\leq& R_{m,n} - \frac{m}{2} \log \frac{ne}{m} \nonumber \\
                                    &\leq& m \log (1 + \sqrt{\frac{m}{n}}),  \label{R.small.m}
\end{eqnarray}
where $ d_1 = O \left( (\frac{m}{n})^{1/3} \right) $.

When $n=o(m)$, the $R_{m,n}$ is near $\textstyle n\log\frac{m}{ne}$ in the following sense.
\begin{eqnarray}
m \log \left( 1 + (1-d_2) \frac{n}{m} \right)\!&\leq&\!R_{m,n} - n\log \frac{m}{ne}  \nonumber  \\
                                                                 \!&\leq&\!m \log \left( 1 + \frac{n}{m} + d_3 \right) \label{R.large.m}
\end{eqnarray}
where $ d_2 = O(\frac{n}{m}) $, and $d_3 = \frac{1}{2\sqrt{\pi}} \frac{n^2e^2}{m(m-ne)}$.

When $n=bm$, the $R_{m,n} = c m$, where the constant $c = a^*b \log e + \log C_{a^*}$, and $a^*$ is such that  $\mathbf E_{P_a} N_1 = b$.
\end{thm}

\begin{IEEEproof}
The expression of the regret is from the definition. The fact that $a^*$ is the minimizer can be seen by taking partial derivative with respect to $a$ of expression (\ref{R.n}). The upper bounds are derived by applying Lemma \ref{lemma1} in the appendix. Pick $a= m/2n$ and use the first inequality, we get the upper bound for $m=o(n)$ case; pick $a = \ln (m/ne)$ and use the second inequality, we have the upper bound for $n=o(m)$. Here $\ln$ is the logarithm base $e$. The rest of the proof is left in Appendix B.
\end{IEEEproof}

\textbf{Remark 1}:
The regret depends only on the number of parameters $m$, the total counts $n$ and the tilting parameter $a$. The optimal tilting parameter is given by a simple moment condition in equation (\ref{m.a}).

\textbf{Remark 2}:
The regret $R_{m,n}$ is close to the minimax level in all three cases listed in Theorem \ref{R}. The main terms in the $m=o(n)$ and $n=o(m)$ cases are the same as the minimax regret given in \cite{szpankowski11} except the multiplier for $\log (ne/m)$ here is $m/2$ instead of $(m-1)/2$ for the small $m$ scenario. For the $n=bm$ case, the $R_{m,n}$ is close to the minimax regret in \cite{szpankowski11} numerically.

\textbf{Remark 3}:
In fact, the regret provides an upper bound for the regret. Recall that
\begin{eqnarray}
\mathbf E_{P_{\underline \lambda}} \log \frac{P_{\underline \lambda}}{Q_a} &\leq& \mathbf E_{P_{\underline \lambda}} \max_{\underline \lambda } \log \frac{P_{\underline \lambda}}{Q_a} \nonumber \\
                                                                                                                        &=& a\lambda_{sum} \log e + m \log C_a. \label{r.unif.ub}
\end{eqnarray}

%However, the performance of independent tilted distributions as a universal compression method under redundancy depends on the actual value of the parameters $\underline \lambda$. 
Theorem \ref{r} in Appendix C gives more detailed expression of the redundancy for using $Q_a$. While there is a reduction of $(m/2) \log e$ bits as compared to the pointwise case, the error depends on the $\lambda_j$'s. Nevertheless, expression (\ref{r.unif.ub}) still provides an uniform upper bound for the redundancy for all possible Poisson means $\underline \lambda$ with a given sum.

\newtheorem{cor}{Corollary}

\begin{cor}
\label{R.mul}
Let $\mathcal P_{\Theta}^{m}$ be a family of multinomial distributions with total count $n$. Then the maximized regret $R(Q_a, \mathcal P_{\Theta}^{m}, S_{m,n})$ has an upper bound within $\textstyle \frac{1}{2} \log 2\pi n$ above the upper bound in Theorem \ref{R}.
\begin{IEEEproof}
This can be easily seen by equation (\ref{N.given}).
\end{IEEEproof}
\end{cor}

\subsection{Subset of sequences with partitioned counts}
\label{subset.sec}

One advantage of using the tilted Stirling ratio distributions is the flexibility of choosing tilting parameters. As mentioned in the introduction, the ratio $m/n$ uniquely determines the optimal tilting parameter. In fact, different tilting parameters can be used for symbols to adjust for their relative importance in the alphabet. Here we consider a situation in which the empirical distribution has most probability captured by a small portion of the symbols. This happens when the sorted probability list is quite skewed. 

The following theorem holds for strings with constraints on the sum of tail counts $\sum_{j>L} N_j = nf$. Small remainder occurs in the following regret bound when $nf/(m-L)$ and $L/(n-nf)$ are both small.

\begin{thm}
\label{subsets}
Let $S_{m,n,f,L}$ be a subset of count vectors with the tail sum controlled by a value $0 \leq f \leq 1$, that is, 
$S_{m,n,f,L}=\{\underline N\!=\!(N_1,\ldots,N_m)\!:\,\sum_{j=1}^m\!N_j\!=\!n,\, \sum_{j>L}\!N_j\!=\!nf \}. $
Here $L$ is a number between $0$ and $m$. The regret of using the tilted Stirling ratio distributions for count vectors in $S_{m,n,f,L}$ given each $ L \in \{0,\ldots, m\} $ is mainly 
\begin{equation}
\frac{L}{2} \log \frac{(n-nf)e}{L} + nf \log \frac{(m-L)}{nf e}. \label{subset} 
\end{equation}
The remainder is bounded below by $r_1$ and above by $r_2$, where 
$$ r_1 =  - d_1 \frac{L}{2} \log e  + (m-L) \log \left (1 + (1-d_2) \frac{nf}{m-L} \right), $$
and
\begin{eqnarray}
r_2 &=& (m-L) \log \left( 1 + \frac{nf}{m-L} + d_3 \right) \nonumber \\
        &&+  L \log \left( 1 + \sqrt{\frac{L}{n-nf}} \right). \nonumber
\end{eqnarray}
Here $d_1$ is $O\left( \left(\frac{L}{n-nf} \right)^{1/3} \right)$ and $d_2$ is $ O \left( \frac{nf}{m-L} \right) $ and $d_3 =  \frac{1}{2 \sqrt{\pi}} \frac{(n f e)^2}{ (m-L)\left( (m-L) - nf e \right)}$.
\end{thm}

\begin{IEEEproof}
Consider the product distribution,
\begin{eqnarray}
Q_{a,b} (\underline N)&=& \prod_{j=1}^m P_{a, b} (N_j)  \nonumber \\
                                   &=& \prod_{j=1}^m \frac{N_j^{N_j} e^{- N_j}}{ N_j!} \frac{e^{-a N_j} e^{-b N_j \mathbf{1}_{\{j>L\}}}}{C_{a,b,j}}, \nonumber
\end{eqnarray}
where $C_{a,b,j} = C_a$ if $ j \leq L$, and $ C_{a,b,j} = C_{a,b}$ is defined as $\sum_{k=0}^\infty k^{k} e^{-(1+a+b)k} / k! $ if $ j > L$. It is in fact using an $L$ dimensional product distribution $Q_a$ on the first $L$ symbols, and an $m-L$ dimensional product distribution $Q_{a+b}$ on the rest.

The regret is the same for any $\underline N \in S_{m,n,f,L}$ given $a$ and $b$. That is,
\begin{eqnarray}
&& R (Q_{a,b}, \mathcal P_{\Lambda}^m, S_{m,n,f,L}) \nonumber \\
&=& na\!\log e + L\log\!C_{a} + nfb\log e + (m\!-\!L)\!\log C_{a,b} \nonumber \\
&=& R(Q_a, \mathcal P_{\Lambda}^L, S_{L, n-nf})\!+\!R(Q_{a+b}, \mathcal P_{\Lambda}^{m-L}, S_{m-L, nf}). \nonumber
\end{eqnarray}
Here $\mathcal P_{\Lambda}^j$ denotes the class of $j$ independent Poisson distributions and $S_{j, k}$ is the set of $j$ independent Poisson counts with sum equal to $k$. In the above case, $j=L$ or $m-L$, and $k=n-nf$ or $nf$.

The choice of $a, b$ providing minimization of $R (Q_{a,b}, \mathcal P_{\Lambda}^m, S_{m,n,f,L})$ is given by the following conditions
\begin{equation}
\begin{array}{l}
\label{cond.a.b}
\displaystyle \mathbf E_{P_{a,b}} \sum_{j=1}^m N_j = n \nonumber \\ \nonumber
\displaystyle \mathbf E_{P_{a,b}} \sum_{j>L} N_j = n f. \nonumber \\
\end{array} 
\end{equation}
This result can be derived by applying Theorem \ref{R} to $R(Q_a, \mathcal P_{\Lambda}^L, S_{L, n-nf})$ and $R(Q_{a+b}, \mathcal P_{\Lambda}^{m-L}, S_{m-L, nf})$  respectively.
\end{IEEEproof}

\textbf{Remark 4}:
The problem here is treated as two separate coding tasks, one for a small alphabet with $L$ symbols having a total count $n-nf$, and the other for a large alphabet with $m-L$ symbols with total count $nf$. The two main terms in expression (\ref{subset}) represent regret from coding the two subsets of symbols, with one set containing $L$ symbols having relatively large counts, and each symbol induces $\textstyle \frac{1}{2} \log \frac{n(1-f)e}{L} $ bits of regret, and the other containing the rest $m-L$ symbols with small counts and together cost $\textstyle nf \log \frac{m}{nfe} $ extra bits. 

\textbf{Remark 5}:
One can arrange more flexibility in what the code can achieve by adding small additional pieces to the code. One is to adapt the choice of $L$ between $0$ and $m$, including $\log(m+1)$ more bits for the description of $L$. Next one can either work with the counts in the given order, or use an additional $\textstyle \log {m \choose L}$ bits to describe the subset that has the $L$ largest counts. Then one uses $\log 1/Q_{a,b}( \underline N )$ bits to describe the counts. Rather than fixing $f$, one works with the empirical tail fraction $\hat f (L)$, where $n \hat f (L)$ is the sum of the counts for the remaining $m-L$ symbols. Finally one has to adapt the choices of $a$ and $b$.  A suggested method of doing so is described in Section \ref{no.n}, in which the $Q_{a,b}$ above is replaced by a mixture over a range of choices of $a$ and $b$.

\subsection{Envelope class}
\label{envelope.sec}
Besides a subset of strings, we can also consider subclass of distributions. Here we follow the definition of envelope class in \cite{boucheron08}. Suppose $\mathcal{P}_{m, f}$ is a class of distributions on ${1,\ldots,m}$ with the symbol probability bounded above by an envelope function $f$, i.e.
$$ \mathcal{P}_{m,f} = \{ P_{\mathbf {\theta}} : \theta_j \leq f(j), j = 1,\ldots,m \}. $$
Given the string length $n$, we know the count of each symbol follows a Poisson distribution with mean $\lambda_j = n \theta_j$, $j=1,\ldots,m$. This transfers an envelope condition from the multinomial distribution to a Poisson distribution, the mean for which is restricted to the following set
$$ \Lambda_{m,f} = \{ \underline \lambda : \lambda_j \leq n f(j), j=1,\ldots,m \}. $$

\begin{thm}
\label{envelope}
The minimax regret of the Poisson class $\Lambda_{m,f} $ with envelope function $f$ has the following upper bound
\begin{eqnarray}
&& R(Q_a, \Lambda_{m,f}, \underline N)  \nonumber \\
&\leq& \min_{L \in \{ 1,\ldots m \}} \frac{L}{2}\!\log\!\frac{n(1\!-\!\bar F(L))}{L} + n\bar F(L)\!\log e + r_3, \nonumber
\end{eqnarray}
where $\bar F(L) = \sum_{j>L} f(j)$, and
\begin{eqnarray}
r_3 = \frac{L}{2(1\!-\!\bar F(L))}\log e + L\log\left(\!1\!+\sqrt{\frac{L}{n(1\!-\!\bar F(L))}}\right). \nonumber
\end{eqnarray}
\end{thm}
\begin{IEEEproof}
A tilted distribution with $a\!=\!L/2n(1\!-\!\bar F(L))$ will give the result. Details are left in Appendix \ref{pf.envelope}.
\end{IEEEproof}

\textbf{Remark 6}:
Here in order for $r_3$ to be small, the tail sum of the envelope function $\bar F(L)$ needs to be small, although the upper bound holds for general envelope function $f$ and $L$. This result is of the same order as the upper bound $ \inf_{L: L \leq n} \left( (L-1)/2 \log n + n \bar F(L) \log e \right) + 2  $ given in \cite{boucheron08}. The first main term in the bound given in Theorem \ref{envelope} also matches the minimax regret given in \cite{xie97} for an alphabet with $L$ symbols and $n(1-\bar F(L))$ data points by Stirling's approximation, i.e.,
\begin{eqnarray}
&& \frac{L-1}{2} \log \frac{n(1-\bar F(L))}{2\pi} + \log \frac{\Gamma{(1/2)}^L}{\Gamma{(L/2)}} \nonumber \\
&\approx& \frac{L-1}{2} \log \frac{n(1-\bar F(L))e}{L} + \frac{1}{2} \log \frac{e}{2}. \nonumber
\end{eqnarray}
The extra $(1/2) \log (n(1-f)e/L)$ is because the tilted distribution allows $m$ free parameters instead of $m-1$. 

\textbf{Remark 7}:
The best choice of tilting parameters for envelope class only depends on the envelope function and the number of symbols $L$ constituting the `frequent' subset. Unlike the subset of strings case discussed before, neither the order of the counts nor which symbols are those with largest counts matters, all we need is an envelope function decaying fast enough when the symbol probabilities are arranged in decreasing order so that $L$ and $\bar F(L)$ are both small.

\subsection{Regret with unknown total count}
\label{no.n}

We know that $a^*$ depends on the value of the total count. However, when the total count is not known, we can use a mixture of tilted distributions $Q(\underline N)$.
\begin{eqnarray}
Q (\underline N) &=& \int_{0}^{m/2} Q_a(\underline N) \frac{1}{m/2} da \nonumber \\
                           &=& \int_{0}^{m/2} \prod_{j=1}^m \frac{N_j^{N _j} e^{-N_j}}{N_j! \, C_a} e^{-a N_j} \frac{2}{m} da \nonumber \\
                           &\leq& M(\underline N) \frac{2}{m} \int_{0}^{\infty} e^{-a N} C_a^{-m} da \nonumber
\end{eqnarray}
Here the upper end of the integrated area is due to inequality (\ref{upper.bound.a}). We have $a^* \leq m/(2n) \leq m/2$.

For any realized non-negative total count $N=k$,  the integrand is maximized at $a^*_k$, defined as solution to the equation $\mathbf E_{P_a} N = k$. And the integral can be approximated by Laplace method,
\begin{eqnarray}
Q (\underline N) &\approx& \frac{2}{m} \left( \prod_{j=1}^m \frac{N_j^{N _j} e^{-N_j}}{N_j!} \right) e^{-a^*_k k} C_{a^*_k}^{-m} \sqrt{\frac{2\pi}{c}}, \nonumber
\end{eqnarray}
where $c= -\frac{\partial ^2}{\partial a^2} \ln \left( e^{-ak} C_a^{-m} \right)|_{a=a^*_k}$. 

Hence the regret induced by $Q(\underline N)$ is
\begin{eqnarray}
\log \frac{M(\underline N)}{Q(\underline N)} &\approx& \log e^{a^*_k k} C_{a^*_k}^{m} + \log \sqrt{\frac{c}{2 \pi}} + \log \frac{m}{2}. \nonumber
\end{eqnarray}
By definition,
\begin{eqnarray}
c &=& m \left. \left( \frac{\frac{\partial^2 C_a}{\partial a^2}}{C_a} - \left(\frac{\frac{\partial C_a}{\partial a}}{C_a} \right)^2 \right)\right|_{a=a^*_k} \nonumber \\
   &\leq& m \left(\left. \frac{\frac{\partial^2 C_a}{\partial a^2}}{C_a}\right) \right|_{a=a^*_k}.  \nonumber
\end{eqnarray}
A similar argument as in the proof of Lemma \ref{expectation} yields an upper bound for the first term
\begin{eqnarray}
\frac{\frac{\partial^2 C_a}{\partial a^2}}{C_a} &\leq& \frac{3}{(2a)^2} + \frac{1}{\sqrt{2\pi}}\left(\frac{3}{e}\right)^{3/2}\frac{1}{2a} \nonumber \\
                                                                       &\leq& 3 C_a^4 +\frac{1}{\sqrt{2\pi}}\left(\frac{3}{e}\right)^{3/2} C_a^2 \nonumber \\
                                                                       &<& \frac{7}{2} C_a^4. \nonumber 
\end{eqnarray}

The second last inequality is by Lemma \ref{lemma1}. 

Hence, we have an upper bound for the regret
\begin{eqnarray}
&& \log e^{a^*_k k} C_{a^*_k}^{m} + \log \frac{m}{2} + \frac{1}{2}\log \left( \frac{7m}{4\pi} C_{a^*_k}^4 \right) \nonumber  \\
&<& \log e^{a^*_k k} C_{a^*_k}^{m} + \frac{3}{2} \log m + 2 \log C_{a^*_k}. \nonumber
\end{eqnarray}
Thus, the extra regret above the optimal level by using $Q (\underline N)$ is approximately no more than $\textstyle \frac{3}{2} \log m + 2 \log C_{a^*_k}$ bits.

Similar argument can show that averaging over the two parameters tilting distribution $Q_{a,b}$ can lead to a distribution that achieves regret not much larger than the minimizing value if the actual total count and tail sum were known beforehand.

\subsection{Conditional distributions induced by $Q_a (\underline N)$}
To account for strings of arbitrary length, our coding strategy $Q_a$ assigns a probability distribution to all finite length strings on $\mathcal A^m$. However, when considering strings of a known length, we are interested to see what the distribution looks like conditioning on a particular number $n$. 

Let $\underline N^n$ denote any count vector in $S_{m,n}$, and $N_x^n$ denote the $x$'s component of $\underline N^n$, where $x \in \{1,\ldots,m\}$. Also, let $M_{mul}$ be the $multinomial(n, \underline \theta)$ maximized likelihood. We have
\begin{eqnarray}
Q_a (\underline N^n | N=n) = \frac{Q_a (\underline N^n)}{Q_a(S_{m,n})} = \frac{M_{mul}(\underline N^n)}{M_{mul}(S_{m,n})}. \label{cond.n}
\end{eqnarray}
The conditioning of $Q_a$ in expression (\ref{cond.n}) reduces the Poisson maximized likelihood (conditioned on the sum $N=n$) to be the same as the multinomial maximized likelihood normalized as indicated, which is indeed the Shtarkov NML distribution for the multinomial family of distributions of counts. 

This conditional distribution of counts, when multiplied by the uniform distribution of strings given the counts, induces a distribution on the strings, i.e.,
$$ P_n (\underline X^n) = P_{unif}(\underline X^n | \underline N^n) Q_a (\underline N^n | N\!=\!n), $$
where $\underline X^n$ is the vector $X_1,\ldots,X_n$.

This sequence of distributions $P_n$ are not compatible and hence do not have extensions to a stochastic process.
To see this incompatibility one looks at the sum 
$$ \sum_{x \in \mathcal A} P_{n+1}(X_1, \ldots, X_n, X_{n+1} = x) $$
and confirms that it is not equal to $P_n(X^n)$. This property is what is called the horizon dependence of NML \cite{bartlett13}.

\subsection{Prediction}
\label{prediction}

A sequence of conditional distributions for $X_{i+1}$ given the past observations $X_1,\ldots,X_i$ for $i<n$ provides a sequential prediction with cumulative log loss defined by $\sum_{i<n} \log 1/P(X_{i+1}|X_1,\ldots,X_i)$. 

There are two natural ways of providing this sequence of conditionals. One is to get the conditionals from the full joint distribution $P_n$, which is horizon dependent as mentioned above. It produces cumulative log loss prediction regret precisely the same as the regret of using $Q_a$ for data compression. The other is by using the sequence of distributions $P_{i+1} (X_1, \ldots, X_{i+1}), i<n$, called sequential NML \cite{teemu08}. The sequential prediction distribution $P_{i+1} (X_{i+1} = x | X_1, \ldots, X_i)$ is proportional to $P_{i+1} ( X_1, \ldots,X_i, X_i+1 = x)$ and accordingly simplifies to
$$ P(X_{i+1}\!=\!x | X_1,\!\ldots\!,X_i) = \frac{ (N_x^i+1)^{N_x^i+1} / {N_x^i}^{N_x^i} }{ \sum_{\tilde x = 1}^m (N_{\tilde x}^i+1)^{N_{\tilde x}^i+1} / {N_{\tilde x}^i}^{N_{\tilde x}^i} }. $$

Note that the prediction rule does not involve $a$. Previous study by Shtarkov\cite{shtarkov87} shows that it is approximately proportional for large $N_x$ to the $N_x + 1/2$ rule of the Jeffreys $Beta(1/2,1/2)$ mixture (also called the Krichevski-Trofimov rule) . Yet it differs importantly from the Jeffreys rule for small counts $N_x$.

However, when using two tilting parameters to adjust for relative importance of symbols within an alphabet, for example, $Q_{a,b}$ in Section \ref{subset.sec}, the predictive distribution does depend on $b$, i.e.,
\begin{eqnarray}
&&\!P(X_{i+1}\!=\!x | X_1,\!\ldots\!,X_i) \nonumber \\
&=& \frac{ e^{-\mathbf 1_{\{x>L\}} b} (N_x^i+1)^{N_x^i+1} / {N_x^i}^{N_x^i} }{ \sum_{\tilde x = 1}^m e^{-\mathbf 1_{\{\tilde x>L\}} b} (N_{\tilde x}^i+1)^{N_{\tilde x}^i+1} / {N_{\tilde x}^i}^{N_{\tilde x}^i} }. \nonumber
\end{eqnarray}
Hence, all symbols beyond $L$ are discounted by an extra fact of $\textstyle e^{-b}$ when predicted by this rule.

\section{Application}
\subsection{Simulation}
\begin{figure}
\begin{center}
\includegraphics[width=2.5in]{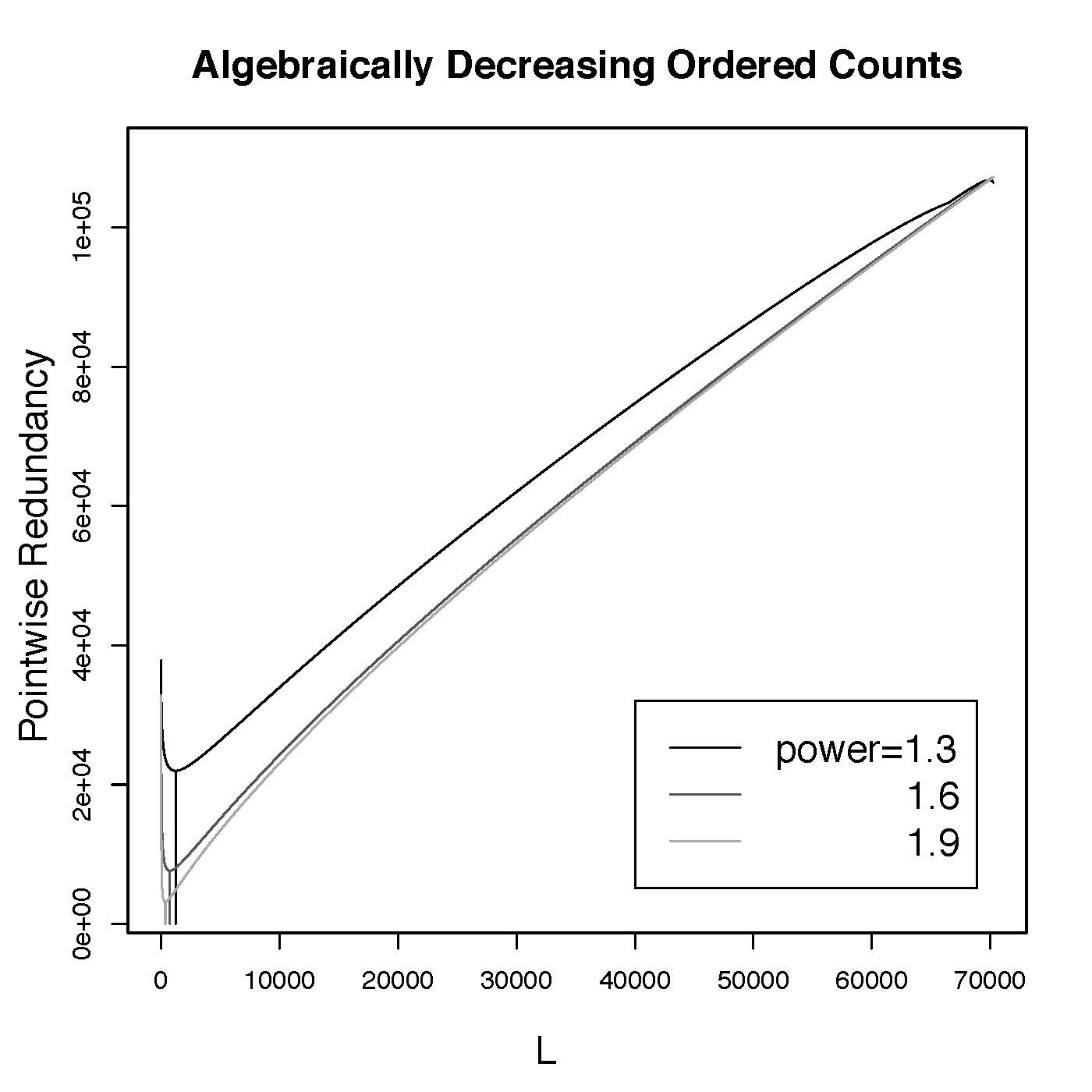}
\caption{Rregretegret of using tilted Stirling ratio distribution for algebraically decreasing counts.}
\label{r.a.sim}
\end{center}
\end{figure}

We first look at the performance of the tilted Stirling ratio distribution for algebraically decreasing counts with simulated data. The alphabet is partitioned into two subsets -- the frequent symbols and the infrequent ones. The tilting parameter is chosen approximately according to the ratio of the number of symbols in a subset and their total count. The regret of assigning different number of symbols as `frequent' ($L$) is shown in Fig. \ref{r.a.sim}.

We can see that more skewness pushes the optimizing $L$ smaller.

\subsection{Real data}

We also provide an example of using the tilted Stirling ratio distribution to code Chinese literature. The target book is an ancient collection of poems named \begin{CJK*}{UTF8}{gbsn}《诗经》\end{CJK*}, translated as the Classic of Poetry. It is the existing earliest collection of Chinese poetry and dates from the 10th to 7th centuries BC \cite{shijing}. The book is downloaded freely from \url{http://wenku.baidu.com/}. Since many ancient words are rarely used today, the encoding is done in GB18030 \cite{gb18030}, the largest Chinese coded character set. It contains $70244$ characters, among which $2889$ appear in the book with a total character count $39161$. There are $792$ characters appear once and $479$ appear twice. The smallest regret happens at $L=2889$ which is the total number of characters appear.

\begin{figure}
\begin{center}
\includegraphics[width=2.5in]{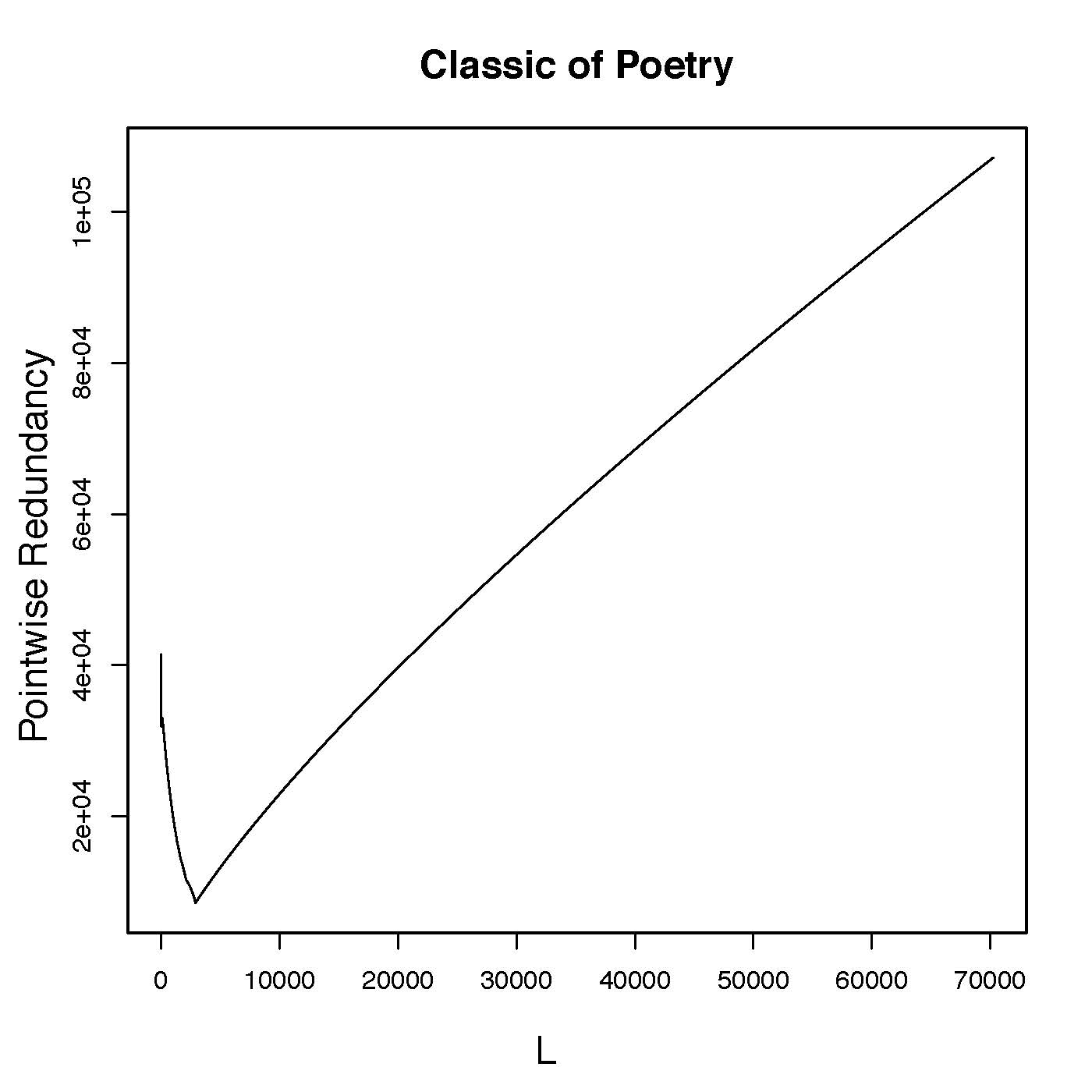}
\caption{Rregretegret of $Q_{a,b}$ for $L$ from $1$ to $m$.}
\label{r.a.pic}
\end{center}
\end{figure}

\section{Discussion}

We have introduced the use of independent tilted maximized Poisson likelihood distributions (also here called tilted Stirling ratio distributions) $Q_a$ for coding the counts for independent random variables. The performance of the coding distribution is close to the minimax level. Actually, the difference between the regret and the minimax level is the probability assigned to the set with the observed total count by the tilted distribution with the optimal tilting parameter, i.e.
\begin{eqnarray}
R(M_{cond}, \mathcal P_{\Lambda}^m, S_{m,n}) &=& R(Q_{a^*}, \mathcal P_{\Lambda}^m, S_{m,n}) \nonumber \\
                                                                                && + \log Q_{a^*} (S_{m,n}). \nonumber 
\end{eqnarray}

The optimal tilting parameter $a^*$ minimizes the difference among all possible $a$. Since $M_{cond}$ reproduces the Shtarkov NML distribution for the multinomial family of distributions on counts, it is the exact pointwise minimax strategy. As shown in this paper, our findings about the regret produced by the distribution $Q_a$, taken together with earlier work \cite{shtarkov87}\cite{xie97}\cite{orlitsky04.2}\cite{szpankowski11}, show that the difference is no larger than about $\log n$ in small alphabet cases, and about $\textstyle \frac{1}{2} \log n$ for moderate or large alphabets. The probability $Q_a(S_{m,n})$ is the probability distribution for the total count $N$ evaluated at $N=n$ as induced by our distribution $Q_a$. Further analysis could be done to characterize this distribution of the total count more precisely.

\appendices

\section{}

\newtheorem{fact}{Fact}

\begin{fact}
\label{fact2}
For any $a>0$,
$$ \frac{1}{\sqrt{2\pi}} \int_0^1 t^{-\frac{1}{2}} e^{-at} dt < \sqrt{\frac{2}{\pi}}. $$
\end{fact}

\begin{IEEEproof}
\begin{eqnarray}
\frac{1}{\sqrt{2\pi}} \int_0^1 t^{-\frac{1}{2}} e^{-at} dt &\stackrel{u=at}=& \frac{1}{\sqrt{2\pi}} \int_0^a (\frac{u}{a})^{-\frac{1}{2}} e^{-u} \frac{1}{a} du \nonumber \\
                                                                                   &=& \frac{1}{\sqrt{2\pi a}} \int_0^a u^{-\frac{1}{2}} e^{-u} du \nonumber
\end{eqnarray}
The integrand is smaller than $u^{-\frac{1}{2}}$ on $[0,a]$, so the integral is upper bounded by 
$$\frac{1}{\sqrt{2\pi a}} \int_0^a u^{-\frac{1}{2}} du = \sqrt{\frac{2}{\pi}}. $$ 
\end{IEEEproof}

\begin{fact}
\label{fact1}
For any $a>0$, 
$$ \sum_{k=1}^\infty \frac{k^{-\frac{1}{2}}}{\sqrt{2\pi} e^{r_k}} e^{-ak} \geq \frac{1}{\sqrt{2\pi}} \int_1^\infty t^{-\frac{1}{2}} e^{-at} dt $$
when $ \frac{1}{12k+1} \leq r_k \leq \frac{1}{12k} $.
\end{fact}

\begin{figure}
\begin{center}
\includegraphics[width=2.5in]{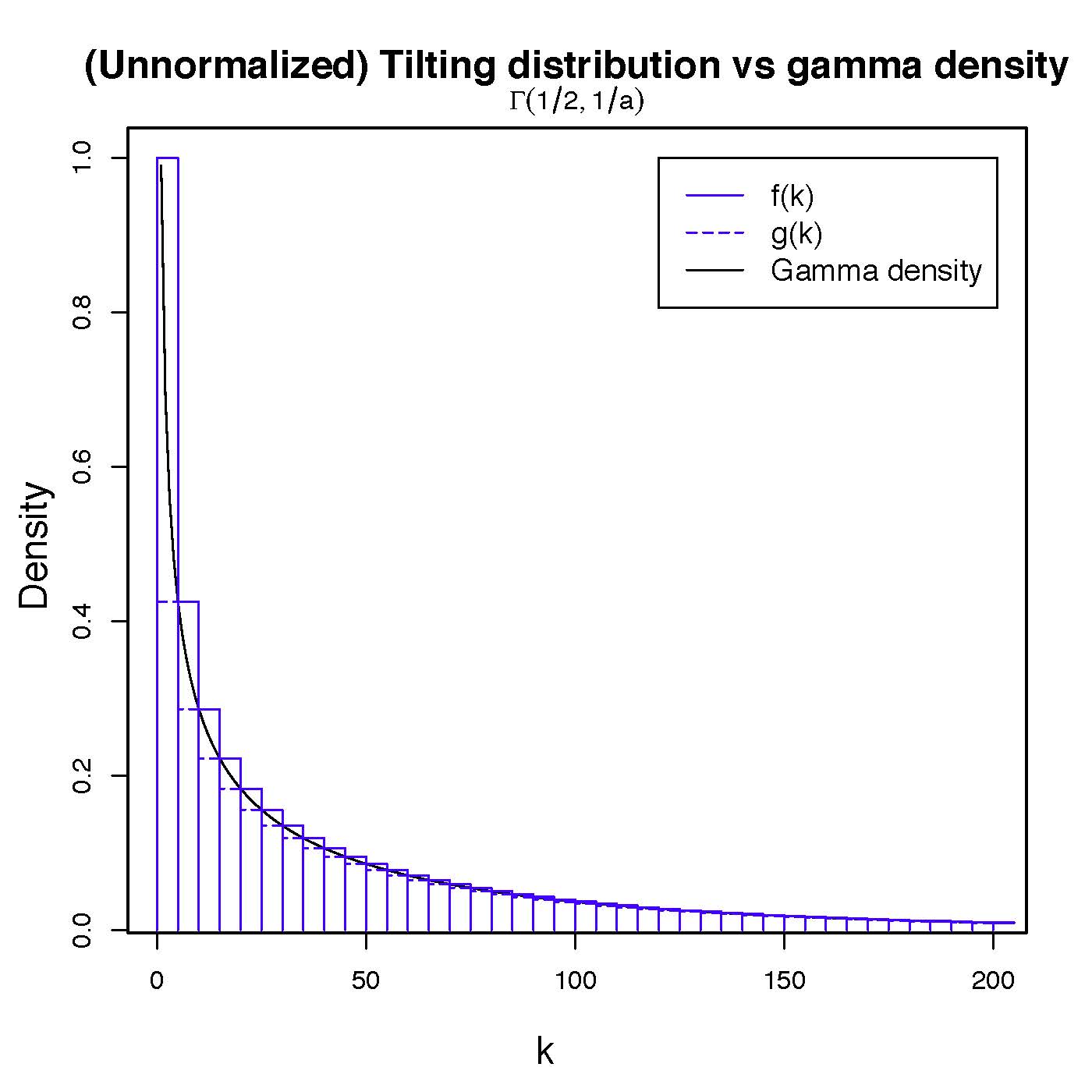}
\caption{tilted distribution vs $\Gamma(\frac{1}{2}, \frac{1}{a})$ density.}
\label{lower.bound.C_a}
\end{center}
\end{figure}

\begin{IEEEproof}
It suffice to show
\begin{equation}
\label{ineq.1}
\sum_{k=1}^\infty \frac{k^{-\frac{1}{2}}}{ e^{\frac{1}{12k}}} e^{-ak} \geq \int_1^\infty t^{-\frac{1}{2}} e^{-at} dt
\end{equation}
As demonstrated in Fig. \ref{lower.bound.C_a}, the curve is bounded between the two step functions representing the two functions $ \textstyle f(k) = k^{-1/2} e^{-ak} $ (the solid line), and $ \textstyle g(k) = (k+1)^{-1/2} e^{-a(k+1)} $ (the dashed line). Note that $ t^{-\frac{1}{2}} e^{-at} $ is convex in $t$, hence the area above the curve is larger than the area below the curve in the rectangles between the two step functions. Although the (unnormalized) tilting probability is shrank by an extra factor of $e^{\frac{1}{12k}}$, as long as it does not drag the step function down below the mid-point of the rectangle, inequality (\ref{ineq.1}) still stands. It remains to show
$$ \frac{e^{\frac{1}{12k}}-1}{e^{\frac{1}{12k}}} \leq  \frac{1}{2} \left( \frac{k^{-\frac{1}{2}} e^{-ak} - (k+1)^{-\frac{1}{2}} e^{-a(k+1)}}{ k^{-\frac{1}{2}} e^{-ak}}  \right), $$
where the left hand side is the part dragged down by the term $e^{\frac{1}{12k}}$ as a portion of the solid line step function, and the right hand side is half of the rectangle between the two step functions as a portion of the same step function. Rearranging the terms and we actually have the following inequality holds for each $k \geq 1$ and $a>0$,
$$ \left( 1 + \left(\frac{k}{k+1}\right)^{\frac{1}{2}} e^{-a} \right) e^{\frac{1}{12k}} \leq 2. $$
Therefore Inequality (\ref{ineq.1}) follows.
\end{IEEEproof}

\newtheorem{lemma}{Lemma}
\begin{lemma}[Bounds for $C_a$]
\label{lemma1}
For any $a>0$, the following bounds hold for $C_a$
\begin{equation}
\max(1, 1 - \sqrt{\frac{2}{\pi}} + \frac{1}{\sqrt{2a}}) < C_a < 1 + \frac{1}{\sqrt{2a}}, \label{ub.small.a}
\end{equation}
and
\begin{equation}
1 + e^{-(a+1)} < C_a < 1 + e^{-(a+1)} + \frac{1}{2 \sqrt{\pi}} \frac{e^{-2a}}{1-e^{-a}}. \label{ub.large.a}
\end{equation}
\end{lemma}

\begin{IEEEproof}
For the upper bound,
\begin{eqnarray}
C_{a} = \sum_{k=0}^\infty \frac{k^k e^{-k}}{k!} e^{-ak} \stackrel{(a)}= 1 + \sum_{k=1}^\infty \frac{k^{-\frac{1}{2}}}{\sqrt{2 \pi} e^{r_k}} e^{-ak}  \label{stirling}
\end{eqnarray}
Here $(a)$ is by Robbins' refinement of Stirling's approximation where $\frac{1}{12k+1} < r_k < \frac{1}{12k}$. 

The sum can be bounded by a gamma integral, so
\begin{eqnarray}
C_a &\leq& 1 + \frac{1}{\sqrt{2\pi}} \int_0^\infty t^{-\frac{1}{2}} e^{-ta} dt  \nonumber \\
        &\leq& 1 + \frac{1}{\sqrt{2\pi}} \frac{\Gamma(\frac{1}{2})}{a^{\frac{1}{2}}} \nonumber \\
         &=& 1 + \frac{1}{\sqrt{2a}}.  \nonumber
\end{eqnarray}

Also, following expression (\ref{stirling}), $C_a$ has the following lower bound.
\begin{eqnarray}
C_a &=& 1 + \sum_{k=1}^\infty \frac{k^{-\frac{1}{2}}}{\sqrt{2\pi} e^{r_k}} e^{-ak} \nonumber \\
        &\stackrel{(b)}\geq& 1 - \sqrt{\frac{2}{\pi}} + \sqrt{\frac{2}{\pi}} + \frac{1}{\sqrt{2\pi}} \int_1^\infty t^{-\frac{1}{2}} e^{-at} dt \nonumber \\
        &\stackrel{(c)}>& 1 - \sqrt{\frac{2}{\pi}} + \frac{1}{\sqrt{2\pi}} \int_0^1 t^{-\frac{1}{2}} e^{-at} dt  \nonumber \\
          && + \frac{1}{\sqrt{2\pi}} \int_1^\infty t^{-\frac{1}{2}} e^{-at} dt \nonumber \\
        &=& 1 - \sqrt{\frac{2}{\pi}} + \frac{1}{\sqrt{2\pi}} \int_0^\infty t^{-\frac{1}{2}} e^{-at} dt  \nonumber \\
        &=& 1 - \sqrt{\frac{2}{\pi}} + \frac{1}{\sqrt{2a}}. \nonumber
\end{eqnarray}
Here again $\frac{1}{12k+1} < r_k < \frac{1}{12k}$, and inequality $(b)$ is due to Fact \ref{fact1} and inequality $c$ is by Fact \ref{fact2}.

Note that inequality (\ref{ub.small.a}) is good for small a. For a moderately large ($a>0.2$), the following upper bound is better.
\begin{eqnarray}
C_{a} &\leq& 1 + e^{-(a+1)} + \sum_{k=2}^\infty \frac{1}{\sqrt{2 \pi k}} e^{-ka}  \nonumber \\
          &<& 1 + e^{-(a+1)} + \frac{1}{2 \sqrt{\pi}} \frac{e^{-2a}}{1-e^{-a}}. \nonumber
\end{eqnarray}
\end{IEEEproof}

\begin{lemma}
\label{expectation}
For any $a>0$,
$$ e^{-(a+1)} \leq \mathbf E_{P_a} N_1 \leq \frac{1}{2a}. $$
\end{lemma}

\begin{IEEEproof}
Let $k^* = \min_{k \in \mathbf N_+} \left| k - \frac{1}{2a} \right|$. We prove the upper bound by consider $a$ within two different intervals. First, if $a < e (\sqrt \pi - \sqrt 2)^2$, we know 
\begin{eqnarray}
&& \sum_{k=1}^\infty \frac{k^{k+1} e^{-k}}{k!} e^{- ak} \nonumber \\
&=& \sum_{k=1}^{k^*-1} \frac{k^{k+1} e^{-k}}{k!} e^{- ak} + \sum_{k=k^*+1}^\infty \frac{k^{k+1} e^{-k}}{k!} e^{- ak} \nonumber \\
      && + \frac{{k^*}^{k^*+1} e^{-k^*}}{k^*!} e^{- ak^*} \nonumber \\
&\stackrel{(a)}\leq& \sum_{k=1}^{k^*-1} \frac{k^{1/2} e^{- ak}}{\sqrt{2\pi}}  + \sum_{k=k^*+1}^\infty \frac{k^{1/2} e^{- ak}}{\sqrt{2\pi}} \nonumber \\
      && + \frac{{k^*}^{1/2} e^{- ak^*}}{\sqrt{2\pi}}  \label{exp.u.b}
\end{eqnarray}
where $(a)$ is an upper bound by Stirling's approximation.

Both sums in the last expression can be upper bounded by a gamma integral, and ${k^*}^{1/2} e^{- ak^*}$ is no larger than the maximum of the unnormalized $gamma(3/2, 1/a)$ density, which is achieved at $1/(2a)$. Hence, we have the following upper bound for expression ($\ref{exp.u.b}$).
\begin{eqnarray}
&&\!\int_{0}^{k^*}\!\frac{t^{1/2}\!e^{-at}}{\sqrt{2\pi}}\!dt\!+\!\int_{k^*}^{\infty}\!\frac{t^{1/2}\!e^{-at}}{\sqrt{2\pi}}\!dt\!+\!\frac{(\!1/2a\!)^{1/2}\!e^{- 1/2}}{\sqrt{2\pi}} \nonumber \\
&=&\!\frac{\Gamma(3/2)}{a^{3/2}\sqrt{2\pi}} + \frac{(1/2a)^{1/2}}{\sqrt{2\pi e}} \nonumber \\
&=&\!\frac{1}{(2a)^{3/2}} +  \frac{1}{\sqrt{2\pi e}} \frac{1}{(2a)^{1/2}} \nonumber
\end{eqnarray}

Using this upper bound for $C_a$, we could prove an upper bound for the expected value.
\begin{eqnarray}
\mathbf E_{P_a} N_1 &=& \sum_{k=1}^\infty \frac{k^{k+1} e^{-k}}{k! \, C_a} e^{- ak} \nonumber \\
                                   &\stackrel{(b)}\leq& \frac{\frac{1}{(2a)^{3/2}} +  \frac{1}{\sqrt{2\pi e}} \frac{1}{(2a)^{1/2}}}{\frac{1}{(2a)^{1/2}} + 1 - \sqrt{\frac{2}{\pi}}} \nonumber \\
                                   &=& \frac{1}{2a} \underbrace{\left( \frac{\frac{1}{(2a)^{1/2}} + \frac{1}{\sqrt{2\pi e}} (2a)^{1/2}}{\frac{1}{(2a)^{1/2}} + 1 - \sqrt{\frac{2}{\pi}}} \right)}_{(A)} \nonumber 
\end{eqnarray}
The lower bound for the denominator in $(b)$ is attributed to Lemma \ref{lemma1}. A little algebra can show that term $(A)$ is no larger than $1$ when $a$ is restricted to $(0,e (\sqrt \pi - \sqrt 2)^2)$.

If $a > e (\sqrt{\pi} -\sqrt{2})^2$, we have $k^*=1$.
\begin{eqnarray}
&& \sum_{k=1}^\infty \frac{k^{k+1} e^{-k}}{k!} e^{- ak} \nonumber \\
&\leq& \sum_{k=1}^\infty \frac{k^{1/2} e^{-ak}}{\sqrt{2\pi}} \nonumber \\
&\stackrel{(c)}\leq& \frac{1}{\sqrt{2\pi}} \left( \int_{0}^{\infty} t^{1/2} e^{-at} dt + \frac{1}{2} e^{-a} \right)  \nonumber \\
&=& \frac{1}{\sqrt{2\pi}} \left( \frac{\Gamma(3/2)}{a^{3/2}} + \frac{1}{2} e^{-a} \right) \nonumber \\
&=& \frac{1}{(2a)^{3/2}} + \frac{1}{2 \sqrt{2\pi}} e^{-a} \nonumber
\end{eqnarray}
where $(c)$ is because the difference between $\int_{t=0}^1 t^{1/2} e^{-at} dt$ and $e^{-a}$ is less than $\frac{1}{2} e^{-a}$. 

By this upper bound for the numerator and Lemma \ref{lemma1} again,
\begin{eqnarray}
\mathbf E_{P_a} N_1 &\leq& \frac{\frac{1}{(2a)^{3/2}} + \frac{1}{2 \sqrt{2\pi}} e^{-a}}{\frac{1}{(2a)^{1/2}} + 1 - \sqrt{\frac{2}{\pi}}} \nonumber \\
&=& \frac{1}{2a} \underbrace{\left( \frac{\frac{1}{(2a)^{1/2}} + \frac{1}{\sqrt{2\pi}} a e^{-a}}{\frac{1}{(2a)^{1/2}} + 1 - \sqrt{\frac{2}{\pi}}} \right)}_{(B)}. \nonumber 
\end{eqnarray} 
Term $(B)$ is no larger than $1$ because $\textstyle \frac{1}{\sqrt{2\pi}} a e^{-a} \leq 1 - \sqrt{\frac{2}{\pi}}$ for all $a$.

For the lower bound, 
\begin{eqnarray}
\mathbf{E}_{p_a}  N_1  &=& \sum_{k=1}^\infty \frac{k^{k+1} e^{-k}}{k!\, C_a} e^{-ak} \nonumber \\
                                     &=& \frac{e^{-(a+1)} \left( \sum_{k=1}^\infty \frac{k^k e^{-(k-1)}}{ (k-1)!} e^{-a(k-1)} \right) }{C_a} \nonumber \\
                                     &\stackrel{l=k-1}=& \frac{e^{-(a+1)} \left( \sum_{l=0}^\infty \frac{(l+1)^{l+1} e^{-l}}{ l!} e^{-al} \right) }{C_a} \nonumber \\
                                     &=& e^{-(a+1)} \underbrace{ \left( \frac{ \sum_{l=0}^\infty \frac{(l+1)^{l+1} e^{-l}}{ l!} e^{-al} }{ \sum_{k=0}^\infty \frac{k^k e^{-k}}{k!} e^{-ak} } \right) }_{(C)}        \label{ratio.C.a} \\
                                     &\stackrel{(d)}\geq& e^{-(a+1)}   \nonumber
\end{eqnarray}
Here inequality $(d)$ is because term $(C)$ is above $1$. Hence the upper bound is deduced.

\end{IEEEproof}

\section{Proof of Theorem \ref{R}}
\begin{IEEEproof}
It remains to show the two lower bounds in expression (\ref{R.small.m}) and (\ref{R.large.m}). In both cases we need a lower bound for $n a^* \log e + m \log C_{a^*}$, and we do it by lower bounding $a^*$ and $C_{a^*}$, respectively. Let $\tilde a = \frac{m}{2n}$.

\begin{itemize}
\item{Bounds for $a^*$}                                                                                                                    
\end{itemize}

We know $a^*$ is the solution for the following equation.
$$ \mathbf{E}_{P_{a^*}} N_1 = \frac{n}{m} $$
By Lemma \ref{expectation}, we have
\begin{eqnarray}
\frac{1}{2 a^*} &\geq& \frac{n}{m} \nonumber
\end{eqnarray}
That gives
\begin{eqnarray}
a^* &\leq& \frac{m}{2n} = \tilde a \label{upper.bound.a}
\end{eqnarray}
Since $C_a$ is decreasing in $a$, we have
$$ C_{a^*} \geq C_{\tilde a} > \frac{1}{\sqrt{2\tilde a}} = \sqrt{\frac{n}{m}}. $$

For any $j \in \{1,\ldots,m\}$, and $a>0$, we have
\begin{eqnarray}
\mathbf{E}_{P_a} N_1 &=& \sum_{k=1}^{\infty} \frac{k^{k+1} e^{-k}}{k!\, C_a} e^{-ak} \nonumber \\
                                         &\stackrel{(a)}\geq& \frac{\sum_{k=1}^{\infty} \frac{k^{k+1} e^{-k}}{k!} e^{-ak}}{1+\frac{1}{\sqrt{2a}}}  \nonumber \\
                                         &\stackrel{(b)}=& \frac{\sum_{k=1}^{\infty} \frac{k^{\frac{1}{2}}}{\sqrt{2\pi} e^{r_k}} e^{-ak}}{1+\frac{1}{\sqrt{2a}}}  \label{l.b.C.a}
\end{eqnarray}
Here $(a)$ is attributed to inequality (\ref{ub.small.a}), step $(b)$ is by Stirling's approximation, and $\frac{1}{12k+1} < r_k < \frac{1}{12k}$.
Pick $k_1 = a^{-1/3} $, then the numerator of expression (\ref{l.b.C.a}) can be lower bounded by
\begin{eqnarray}
&& \sum_{k=\lfloor k_1 \rfloor}^{\infty} \frac{k^{1/2}}{\sqrt{2\pi} e^{r_k}} e^{-ak} \nonumber \\
&\geq& \sum_{k=\lfloor k_1 \rfloor}^{\infty} \frac{k^{1/2}}{\sqrt{2\pi} e^{\frac{1}{12 \lfloor k_1 \rfloor }}} e^{-ak} \nonumber \\
&\geq& \frac{1}{\sqrt{2\pi} e^{\frac{1}{12 (k_1-1)}}} \int_{\lfloor k_1 \rfloor}^\infty t^{1/2} e^{-at} dt \nonumber 
\end{eqnarray}
Taking the integral from $0$ to $\infty$ and subtracting the part from $0$ to $k_1$ yields the lower bound
\begin{eqnarray}
&& \frac{1}{\sqrt{2\pi} e^{\frac{1}{12 (k_1-1)}}} \left( \frac{\Gamma(3/2)}{a^{3/2}} - \int_0^{k_1} t^{1/2} e^{-at} dt  \right) \nonumber \\
&\geq& \frac{1}{\sqrt{2\pi} e^{\frac{1}{12 (k_1-1)}}} \left( \frac{\Gamma(3/2)}{a^{3/2}} - \int_0^{k_1} t^{1/2} dt   \right) \nonumber \\
&=& \frac{1}{\sqrt{2\pi} e^{\frac{1}{12 ( k_1 -1)}}} \left( \frac{\Gamma(3/2)}{a^{3/2}} - \frac{2}{ 3 a^{1/2}} \right).\nonumber
\end{eqnarray}

Write $ r_a = \frac{1}{12(k_1-1)} = \frac{a^{1/3}}{12 ( 1 - a^{1/3})} $. By the above calculation, we have a lower bound for the expectation under the tilting distribution. For $a^*$,
\begin{eqnarray}
\frac{\frac{1}{\sqrt{2\pi} e^{r_{a^*}}} \left( \frac{\Gamma(3/2)}{ {a^*}^{3/2}} - \frac{2}{3 {a^*}^{1/2}} \right) }{1+\frac{1}{\sqrt{2 a^*}}} \leq \mathbf{E}_{a^*} N_1 = \frac{n}{m}. \nonumber 
\end{eqnarray}

Arranging the terms, we have 
\begin{eqnarray}
\frac{1}{2 a^*} &\leq& \frac{n}{m} \left(1 + \sqrt{2 a^*} \right) e^{r_{a^*}} + \frac{2}{3\sqrt{\pi}}  \nonumber \\
                      &\stackrel{(c)}\leq& \frac{n}{m} \left (1 + \sqrt{2 \tilde a} \right) e^{r_{\tilde a}} +  \frac{2}{3 \sqrt{\pi} } \nonumber
\end{eqnarray}

Here $(c)$ is because $a^* \leq \tilde a$ by inequality (\ref{upper.bound.a}). So,
\begin{eqnarray}
a^*
&\geq& \frac{ \tilde a }{ \left(1 + \sqrt{ 2 \tilde a } \right ) e^{r_{\tilde a}} + \frac{4}{ 3 \sqrt{\pi} } \tilde a  }  \nonumber 
\end{eqnarray}
By Taylor expansion, this is no smaller than
\begin{eqnarray}
&& \frac{ \tilde a }{ \left(1 + \sqrt{ 2 \tilde a } \right ) \left( 1 + r_{\tilde a} + O(r_{\tilde a}^2) \right) + \frac{4}{ 3 \sqrt{\pi} } \tilde a  }  \nonumber  \\
              &=& \tilde a \left( 1 - \frac{  r_{\tilde a} +  \sqrt{ 2 \tilde a } +  \sqrt{ 2 \tilde a } r_{\tilde a} + \frac{4}{ 3 \sqrt{\pi}} \tilde a + O(r_{\tilde a}^2) }{ \left( 1 +  \sqrt{ 2 \tilde a } \right) \left( 1 + r_{\tilde a} + O(r_{\tilde a}^2) \right) + \frac{4}{ 3 \sqrt{\pi} } \tilde a } \right)   \nonumber  \\
              &\geq& \tilde a \left( 1 -   r_{\tilde a} - \sqrt{ 2 \tilde a } -  \sqrt{ 2 \tilde a } r_{\tilde a} - \frac{4}{ 3 \sqrt{\pi}} \tilde a - O(r_{\tilde a}^2) \right)  \nonumber
\end{eqnarray}
When $m=o(n)$, $r_{\tilde a}$ is the leading term, so 
\begin{eqnarray}
a^* \geq \tilde a \left( 1 - O\left( r_{ \tilde a} \right) \right) = \frac{m}{2n} \left( 1 - O\left( \left(\frac{m}{n}\right)^{\frac{1}{3}} \right) \right)   \nonumber
\end{eqnarray}
As a result,
\begin{eqnarray}
\nonumber
n a^* \log e &\geq& \left( 1 - O\left( \left(\frac{m}{n}\right)^{\frac{1}{3}} \right) \right) \frac{m}{2}  \log e \nonumber 
\end{eqnarray}
Hence we get inequality (\ref{R.small.m}).

The above lower bound works when $a^*$ is small (i.e., when $m$ is small compared to $n$), yet when it is large, the following bound is better. Let $a_0 = \ln \frac{m}{ne}$.

For large $m$, $a_0$ is the solution of the following equation.
\begin{equation}
\nonumber
\mathbf{E}_{P_{a}} N_1 = \frac{n}{m} 
\end{equation}
That is
\begin{equation}
\nonumber
\sum_{k=1}^\infty \frac{k^{k+1} e^{-k}}{k!\, C_{a_0}} e^{- a_0 k} = \frac{ n }{m} \\ \nonumber
\end{equation}

The left hand side is lower bounded by $e^{-(a_0 + 1)}$ by Lemma \ref{expectation}. Hence we have
\begin{eqnarray}
\nonumber
e^{-(a^* + 1)} &\leq& \frac{ n }{m} \\ \nonumber
e^{a^* } &\geq& \frac{ m}{n e} = e^{a_0} \\ \nonumber
a^* &\geq& a_0 \nonumber
\end{eqnarray}
Thus,
\begin{equation}
na^* \log e \geq n a_0 \log e = n \log \frac{ m}{n e} \nonumber
\end{equation}

\begin{itemize}

\item{Bounds for $C_{a^*}$}
\end{itemize}

Now we want to lower bound $C_{a^*}$. Recall inequality (\ref{ratio.C.a}), let term $(C)$ be defined as 
$$ r_{a} = \frac{ \sum_{l=0}^\infty (l+1)^{l+1} e^{-l} e^{-al} / l! }{ \sum_{k=0}^\infty k^k e^{-k} e^{-ak} /k! }. $$ 
We have
\begin{equation}
r_{a^*} e^{-(a^* + 1)} = \mathbf{E}_{P_{a^*}}  N_j = \frac{n}{m} =  e^{-(a_0 + 1)}. \nonumber                                                   
\end{equation}
It gives
$$ e^{-(a^* + 1)} = \frac{e^{-( a_0 + 1)}}{r_{a^*} }. $$
By definition,
\begin{equation}
C_{a^*} \geq 1 + e^{-(a^* + 1) } = 1 + \frac{e^{-( a_0 + 1)}}{r_{a^*} }. \nonumber
\end{equation}
Hence, we have
\begin{eqnarray}
\frac{1}{r_{a^*}}  &\geq&  \frac{ 1 + e^{-(a^* + 1)} }{ 1 + 4e^{-(a^* + 1)} + \frac{27}{2} \frac{e^{-2(a^* + 1)}}{1-e^{-(a^* + 1)}} } \nonumber \\ 
                                          &=& 1 - \frac{ 3e^{-(a^* + 1)} + \frac{27}{2} \frac{e^{-2(a^* + 1)}}{1-e^{-(a^* + 1)}} }{ 1 + 4e^{-(a^* + 1)} + \frac{27}{2} \frac{e^{-2(a^* + 1)}}{1-e^{-(a^* + 1)}} }  \nonumber \\
                                          &\geq& 1 - 3e^{-(a^* + 1)} - \frac{27}{2} \frac{e^{-2(a^* + 1)}}{1-e^{-(a^* + 1)}} \nonumber \\
                                          &=&  1 - O \left( e^{-(a^* + 1)} \right) \nonumber \\
                                          &\geq&  1 - O \left( e^{-( a_0 + 1)} \right). \nonumber
\end{eqnarray}
So
\begin{equation}
C_{a^*} \geq 1 + \left(1- O \left( e^{-( a_0 + 1)} \right) \right) e^{-(a_0 + 1)} \nonumber
\end{equation}
And we derive a lower bound
\begin{equation}
m \log C_{a^*} \geq m \log \left( 1 + \left(1 - O \left( \frac{n}{m} \right)  \right) \frac{n}{m} \right). \nonumber 
\end{equation}

Therefore, inequality (\ref{R.large.m}) follows.
\end{IEEEproof}

\section{Redundancy}
\label{r.pf}

\setcounter{thm}{-1}
\begin{thm}
\label{thm0}
Let $M$ denote the Stirling ratio distribution as defined before, and $M_{cond}$ be the measure $M$ conditional on the observed value $ \frac{1}{m} \sum_{j=1}^m h(N_j) = \alpha$, where $h$ is a function of the data, and let $P_a$ be the tilted distribution with parameter $a$, chosen by the condition $\mathbf{E}_{P_a} h(N_1) = \alpha $, and $\mathcal C_{\alpha}$ be a class of distributions with the expected value equal to the observed
$$ \mathcal C_{\alpha} = \{ P : \mathbf{E}_{P} h(N_1) = \alpha \}. $$ 
Similar to what has been shown in \cite{csiszar75}, \cite{csiszar84}, and \cite{campenhout81} for i.i.d random variables, $Q_a=\otimes_{j=1}^m P_a$ is the information projection of $M$ on $\mathcal C_{\alpha}$.
In fact,
\begin{eqnarray}
D(M_{cond} || M) &=& D(M_{cond} || Q_a) + D( Q_a || M) \nonumber \\
                            &\geq& D( Q_a || M). \nonumber
\end{eqnarray}
\end{thm}
Theorem \ref{thm0} says the tilted distribution is closest to the original distribution in relative entropy among all distributions with the expected value of a function equal to the observed value. Hence it is the redundancy minimizing distribution over the class of distributions with a given moment condition. In particular, if $h(x)=x$, the condition is made on the total count.

\setcounter{thm}{+3}
\begin{thm}
\label{r}
Let $X_1,\ldots,X_N$ be generated from an alphabet $\mathcal A$ of size $m$, where $N$ is also random. Let $\lambda_{sum}$ denote the mean of $N$, i.e. $ \lambda_{sum}:= \sum_{j=1}^m \lambda_j $, with $\lambda_{sum}$ known and $\lambda_j > 0 $ for all $j$. Let $\mathcal P^m_{\lambda_{sum}}$ denote the class of distributions on $N_1,\ldots,N_m$ with expected total count equal to $\lambda_{sum}$. The redundancy by using a tilted distribution $Q_a$ is mainly 
$$ (A) = \left( (-\frac{m}{2} + a \lambda_{sum} ) \log e + m\log C_a \right), $$
with the error bounded by
$$ \sum_{j=1}^m (\frac{1}{3\lambda_j^2} + \frac{5}{6\lambda_j} ) \log e. $$
The error is small if no $\lambda_j$ is too small.
Moreover, the minimizer of the redundancy is $a^*$. Therefore $P_{a^*}$ is an minimax strategy within the class of distributions with mean $\lambda_{sum}$. The minimax redundancy has the following approximations according to the magnitude of $m$ and $\lambda_{sum}$.

When $m = o(\lambda_{sum})$, term (A) satisfies the following inequality
\begin{equation}
0 \leq \left| (A) - \frac{m}{2} \log \frac{\lambda_{sum}}{m} \right| \leq m \log (1+ \sqrt{\frac{m}{\lambda_{sum}}}). \label{r.small.m}
\end{equation}

When $\lambda_{sum} = o(m) $, term (A) satisfies the following inequality
\begin{eqnarray}
&& m\log \left(1+\frac{\lambda_{sum}}{m} \right) -\lambda_{sum} \log e \nonumber \\
&\leq& \left| (A) - \left( \lambda_{sum} \log \frac{m}{ \lambda_{sum}}  - \frac{m}{2} \log e \right) \right| \nonumber \\
&\leq& \frac{1}{2\sqrt{\pi}} \frac{\lambda_{sum}^2 e^2}{m-\lambda_{sum} e} \log e.      \label{r.large.m} 
\end{eqnarray}
\end{thm}

\begin{IEEEproof}
The first part of the proof follows Lemma 3 in \cite{xie97}, and the second part resembles proof of Theorem \ref{R}.

\begin{eqnarray}
&& \mathbf{E}_{\underline{\lambda}} \ln \frac{\prod_{j=1}^m P_{\lambda_j} (N_j)}{ Q_a (\underline N) } \nonumber \\
&=& \sum_{j=1}^m \left( \lambda_j \ln \lambda_j \right) - \sum_{j=1}^m \mathbf{E}_{\lambda_j} \left( N_j \ln N_j \right) + a \sum_{j=1}^m \lambda_j \nonumber \\
&&+ m \ln C_a \nonumber
\end{eqnarray}
Following Lemma 3 in \cite{xie97}, by Taylor's expansion, for each $j$,
\begin{eqnarray}
&& \mathbf{E}_{\lambda_j} \left( N_j \ln N_j \right)  \nonumber \\
&\geq& \lambda_j \ln \lambda_j + \mathbf{E}_{\lambda_j} (N_j - \lambda_j) (1 + \ln \lambda_j) \nonumber \\
&& + \mathbf{E}_{\lambda_j} \frac{1}{2} (N_j-\lambda_j)^2 \frac{1}{\lambda_j} + \frac{1}{6} \mathbf{E}_{\lambda_j} (N_j - \lambda_j)^3(-\frac{1}{\lambda_j^2}) \nonumber \\
&=& \lambda_j \ln \lambda_j + \frac{1}{2} - \frac{1}{6\lambda_j}. \nonumber
\end{eqnarray}
We also know by Jensen's Inequality that
$$ \mathbf{E}_{\lambda_j} \left( N_j \ln N_j \right) \geq \lambda_j \ln \lambda_j. $$
Hence,
$$ \mathbf{E}_{\lambda_j} \left( N_j \ln N_j \right) \geq \lambda_j \ln \lambda_j + \frac{1}{2} + \max( - \frac{1}{6\lambda_j}, - \frac{1}{2}). $$
And
\begin{eqnarray}
&& \mathbf{E}_{\lambda_j} \left( N_j \ln N_j \right) \nonumber \\
&\leq& \lambda_j \ln \lambda_j + (\mathbf{E}_{\lambda_j} N_j - \lambda_j)(1+ \ln \lambda_j) \nonumber \\
&& + \frac{\mathbf{E}_{\lambda_j} (N_j - \lambda_j)^2}{2\lambda_j} - \frac{\mathbf{E}_{\lambda_j}(N_j-\lambda_j)^3}{6\lambda_j^2} \nonumber \\
&& + \frac{\mathbf{E}_{\lambda_j}(N_j-\lambda_j)^4}{3\lambda_j^3} \nonumber \\
&=& \lambda_j \ln \lambda_j + \frac{1}{2} + \frac{1}{ 3 \lambda_j^2} + \frac{5}{6\lambda_j}. \nonumber
\end{eqnarray}

Therefore,
\begin{eqnarray}
&& - \left( \sum_{j=1}^m \frac{1}{3\lambda_j^2} +  \frac{5}{6\lambda_j} \right) \nonumber \\
&\leq& \mathbf{E}_{\underline{\lambda}} \ln \frac{\prod_{j=1}^m P_{\lambda_j} (N_j)}{ Q_a (\underline N) } \nonumber \\
&& - \left( - \frac{m}{2} + a \sum_{j=1}^m \lambda_j + m\ln C_a  \right) \nonumber \\
&\leq& \min \left( \sum_{j=1}^m \frac{1}{6\lambda_j}, \frac{m}{2} \right).  \nonumber
\end{eqnarray}

The fact that $a^*$ is the minimizer can be easily seen by taking partial derivative with respect to a for term (A). The two inequalities are attributed to Lemma \ref{lemma1}, by picking $a = m/(2 \lambda_{sum})$ and $a = \ln (m/\lambda_{sum} e)$ respectively.
\end{IEEEproof}

\section{Proof of Theorem \ref{envelope}}
\label{pf.envelope}

\begin{IEEEproof}
The MLE for an envelope class is the following
\begin{equation}
\hat \lambda_j =\arg \sup_{\lambda_j \leq nf(j)} P_{\lambda_j} (N_j) = N_j \wedge n f(j). \nonumber
\end{equation}

We formulate a tilted distribution by multiplying the exponential tilting factor $\textstyle e^{-aN_j}$ for each $j \in \{1,\ldots,m\}$ and normalize it.
\begin{eqnarray}
\nonumber
P_a (N_j) = 
\left\{ 
\begin{array}{rcl}
        \frac{N_j^{N_j} e^{-N_j}}{N_j!} \frac{e^{-a N_j}}{C_{a,j}}  & \mbox{if} & N_j \leq nf(j) \\
        \frac{(nf(j))^{N_j} e^{-nf(j)}}{N_j!} \frac{e^{-a N_j}}{C_{a,j}}  & \mbox{if} & N_j > nf(j) \\
\end{array}\right.
\end{eqnarray}
where $C_{a,j} = \sum_{N_j \leq nf(j)} \frac{N_j^{N_j} e^{-N_j}}{N_j ! } e^{-a N_j} + \sum_{N_j > nf(j)} \frac{(nf(j))^{N_j} e^{-nf(j)}}{N_j ! } e^{-a N_j} $. 

The regret of using independent $P_a$ for each $N_j$ in $\underline N \in S_{m,n}$ is
\begin{equation}
\label{R.envelope}
\log \prod_{j=1}^m \frac{P_{\hat \lambda_j} (N_j)}{P_a (N_j)} = na \log e + \sum_{j=1}^m \log C_{a,j}. 
\end{equation}
Again, $a^*$ minimizes expression (\ref{R.envelope}).

For each $j$ and any positive $a$,
\begin{eqnarray}
C_{a,j} &=& \sum_{N_j \leq \lfloor nf(j) \rfloor } \frac{N_j^{N_j} e^{-N_j}}{N_j !} e^{-a N_j}  \nonumber \\ 
              && + \sum_{N_j>nf(j)} \frac{(nf(j))^{N_j} e^{-nf(j)}}{N_j !} e^{-a N_j}. \nonumber
\end{eqnarray}
The sum only depends on the envelope function $f(j)$ for given $a$ and $j$. 

Since $ (nf(j))^x e^{-nf(j)} \leq x^x e^{-x}$ for all $x>0$,
for any symbol $j$ with $N_j > nf(j)$, we have
$$ \frac{(nf(j))^{N_j} e^{-nf(j)}}{N_j !} e^{-a N_j} \leq \frac{N_j^{N_j} e^{-N_j}}{N_j !} e^{-a N_j}.  $$
Hence we have,
\begin{eqnarray}
C_{a,j} &\leq& \sum_{N_j = 0}^\infty \frac{N_j^{N_j} e^{-N_j}}{N_j !} e^{-a N_j} \nonumber \\
            &\leq& 1 + \sum_{N_j=1}^{\infty} \frac{N_j^{N_j} e^{-N_j}}{N_j !} e^{-a N_j} \nonumber \\
            &\stackrel{(a)}\leq& 1 + \sum_{N_j=1}^{\infty} \frac{1}{\sqrt{2 \pi}} N_j^{-1/2} e^{-a N_j} \nonumber \\
            &\leq& 1 + \frac{1}{\sqrt{2\pi}} \int_{0}^{\infty} t^{-\frac{1}{2}} e^{-at} dt \nonumber \\ 
            &=& 1 + \sqrt{\frac{1}{2a}} \nonumber                                                                                                                                                                                                                                                 
\end{eqnarray}
where $(a)$ is by Stirling's approximation. This is a good bound when $nf(j)$ is big.

However, if $nf(j)$ is small, the following upper bound is better. For $N_j \leq \lfloor nf(j) \rfloor$,
\begin{eqnarray}
\nonumber
\sum_{N_j \leq \lfloor nf(j) \rfloor } \frac{N_j^{N_j} e^{-N_j}}{N_j !} e^{-a N_j} &\leq& \sum_{N_j \leq \lfloor nf(j) \rfloor } \frac{N_j^{N_j}}{N_j!} \nonumber \\
                                                                                                                        &\leq& \sum_{N_j \leq \lfloor nf(j) \rfloor } \frac{\left( nf(j) \right) ^{N_j}}{N_j!}. \nonumber
\end{eqnarray}

For the second partial sum, we also have
\begin{eqnarray}
\nonumber
&&\sum_{N_j>nf(j)} \frac{(nf(j))^{N_j} e^{-nf(j)}}{N_j !} e^{-a N_j} \nonumber \\
&\leq&  \sum_{N_j>nf(j)} \frac{\left(nf(j)\right)^{N_j}}{N_j!}. \nonumber 
\end{eqnarray}

Deduce,
\begin{equation}
C_{a,j} \leq \sum_{N_j=0}^\infty \frac{\left(nf(j)\right)^{N_j}}{N_j!} = e^{nf(j)}. \nonumber
\end{equation}

Hence for any given $a$, $j$ and $L \in \{ 1, 2, \ldots, m\}$, the following upper bound holds.
\begin{eqnarray}
&& na \log e + \sum_{j=1}^m \log C_{a,j} \nonumber \\
&\leq& na \log e \nonumber \\
&& + \log \left( \prod_{j=1}^L \left( 1 + \sqrt{\frac{1}{2a}} \right) \prod_{j = L+1}^m \left( e^{nf(j)} \right) \right) \nonumber \\
&=& na \log e + L \log \left( 1 + \sqrt{\frac{1}{2a}} \right) \nonumber \\
&& + \left ( \sum_{j = L+1}^m nf(j) \right) \log e. \nonumber 
\end{eqnarray}

Let $a = \frac{L}{2 \left(n-\sum_{j>L} nf(j) \right) }$, the result follows.
\end{IEEEproof}

\section*{Acknowledgment}

The authors would like to thank Prof. Mokshay Madiman and Prof. Joseph Chang for their helpful advice and consistent support. They are also grateful to the family-like department and classmates, which makes all of these possible.

\ifCLASSOPTIONcaptionsoff
  \newpage
\fi

\bibliographystyle{IEEEtran}
\bibliography{/users/Littlelittle/Documents/study/Latex/Reference}

\end{document}